\documentstyle[12pt,a4,psfig]{article}
\title
{Quark Mass Hierarchy, FCNC 
and $CP$ violation  in a Seesaw model}

\author{Y. Kiyo$^1$,~ T. Morozumi$^1$,~ P.Parada$^{2,3}$,~
M.N.Rebelo$^{3,4}$ and M.Tanimoto$^{4,5}$}

\date{}

\begin{document}

\maketitle

{\small 
\begin{enumerate}
\item Dept.~of Physics, Hiroshima University, 
739-8526 Higashi-Hiroshima, Japan.
\item Dep.~de F\'\i sica, Universidade
da Beira Interior, Rua Marqu\^es d'\'Avila e Bolama,\\
6200 Covilh\~a, Portugal.
\item CFIF/IST, Av. Rovisco Pais, P-1096 Lisboa Codex, Portugal.
\item Dep.~de F\'\i sica, Instituto Superior T\'ecnico, 
Av. Rovisco Pais, P-1096 Lisboa Codex, Portugal.
\item Science Education Laboratory, Ehime University, 
Bunkyo-cho, 790-8577 Matsuyama, Japan.
\end{enumerate}}
\begin{picture}(5,4)
       \put(360,312){HUPD-9823}
       \put(360,300){GATC-98-1}
       \put(360,288){hep-ph/9809333}
\end{picture}
\thispagestyle{empty}
\def\beq{\begin{equation}}
\def\eeq{\end{equation}}
\def\bea{\begin{eqnarray}}
\def\eea{\end{eqnarray}}
\def\nn{\nonumber}
\begin{abstract}
The seesaw model of quark masses is studied systematically,
focusing on its developments.
A framework allowing the top quark mass to be of the order
of  the electroweak symmetry breaking scale, while the remaining
light quarks have much smaller masses, due to the seesaw mechanism, is
presented. The violation of the GIM mechanism is shown to be
small and the tree level FCNC are suppressed naturally.
In this model, there are many particles which could contribute to
the FCNC in the one-loop level. Parameters of the model are constrained
 by using the experimental data on $K^0-\overline K{}^0$ mixing and
$\epsilon_K$.
The rare K meson decays
$K_{L,S} \rightarrow \pi^{0} \nu \bar{\nu}$ and
$K^{+} \rightarrow \pi^{+} \nu \bar{\nu}$ are also investigated in the
model.
In these processes the scalar operators 
$(\bar{s}d)(\bar{\nu}_{\tau}\nu_{\tau})$,
which are derived from
box diagrams in the model, play an important role due to
an enhancement factor $M_{K}/m_{s}$ in the matrix element
$<\pi|\bar{s}d|K>$. It is  emphasized   that the $K_{L}$
decay process through the scalar operator is not the $CP$ violating
mode, so  $B(K_L \rightarrow \pi^{0}\nu \bar{\nu})$ remains
non-zero even in the $CP$ conserved limit. 
The pion energy spectra
for these processes are predicted.
\end{abstract}
\section{Introduction}
The idea of the seesaw mechanism  for  the neutrino mass \cite{Yana}
\cite{Grs}
was extended to  quark masses
sometime ago \cite{Bere}-\cite{Moha}.
At that time, the top
quark mass was not known and it was not expected to be
as heavy as  the electroweak symmetry breaking scale.
Therefore all charged fermions  were assumed to be lighter 
than $10^2 ~{\rm GeV}$.    The motivation for the
construction of the model was to give an  explanation
for the smallness of the  masses of the
charged fermions (quarks and charged leptons) as compared
to the electroweak breaking scale. However, now we know that
the top quark is as heavy as O($10^2 ~{\rm GeV}$) and that
has to be taken into account.
Koide and Fusaoka (KF) \cite{KF} and T. Morozumi, 
T. Satou, M. N. Rebelo and M. Tanimoto \cite{MORO}
studied the top quark mass problem in a
seesaw model.
Based on this  successful approach, 
we can discuss implications of the seesaw model focusing on the 
experimental data
such as flavor changing neutral currents (FCNC) and $CP$ violation.
In this paper, we present in detail the formalism and make a systematic 
phenomenological analysis of the implications of the model.
A brief introduction to the Dirac seesaw
scheme is given in section 2, in which the incorporation of the top
quark is outlined. In section 3
we show how the mass hierarchy and the flavor
mixing are introduced. Section 4 
is devoted to
estimating tree level FCNC. In section 5,
the flavor mixing in the charged currents is
studied and the parameterization of the generalized
Cabbibo Kobayashi Maskawa (CKM) matrix is specified.
 \cite{Cabbibo} \cite{KM} In section 6,
  the $K^0-\overline K^0$ mixing and the $CP$ violating
parameter $\epsilon_K$ are studied and the constraint on the
parameters of the present model is obtained.  
In section 7,  
the rare K meson decays $K \rightarrow \pi \nu
\bar{\nu}$ are studied in the present model.  
The effect of the scalar interaction 
on the resulting pion energy spectrum is given.
Our conclusions are presented in section 8.
\section{Seesaw Model and Top Quark Mass}
Let us first summarize the
ideas of the seesaw model for quark masses.
The gauge group of the standard model (SM) is extended
to $SU(2)_L \times SU(2)_R \times U(1)$, so that the 
usual Yukawa mass terms for the charged fermions are not
allowed, since now 
the right handed charged fermions belong
to an $SU(2)_R$ doublet.  That is:\\ 
\beq
\quad \delta_{SU(2)_{R}} \left( \bar\psi_L \phi d_R \right)
\ne 0.
\label{eq:1}
\eeq   
The $SU(2)_R$ gauge boson must be much heavier than the 
$SU(2)_L$ gauge boson
because we do not see any deviation from the V-A structure 
of the charged currents, thus the symmetry has to be broken 
at a sufficiently high energy scale. 
The representation of the
Higgs breaking $SU(2)_R$ is chosen in such a way that
no renormalizable mass term for the standard-like fermions
is allowed. Therefore, for example, the 
Higgs bi-doublet $M (2,2)$ under $SU(2)_L \times SU(2)_R$ is
excluded.
A single  Higgs doublet, $\phi_R (1,2)$, is used for
breaking $SU(2)_R$.  With this Higgs field, the possible mass term
for the standard-like quarks is the dimension 5 operator,
\beq
{1\over \Lambda_{NEW}} \bar{\psi_L} \phi_L \phi_R^\dagger
\psi_R,
\label{eq:2}
\eeq 
where $\Lambda_{NEW} $ is some new physics scale. The scale
must be larger than $\eta_R$ (VEV of $\phi_R$).
Quarks and charged leptons acquire
their masses, which are much smaller than
the electroweak breaking scale, $\eta_L$ (i.e., VEV of $\phi_L$),
through the dimension 5 operator:
\beq
m_f= {\eta_L \eta_R \over \Lambda_{NEW}},
\label{eq:3}
\eeq
where ${\eta_R \over \Lambda_{NEW}}<<1.$ 
Now we can think of what is the
fundamental theory behind this dimension 5 operator.
The interaction can be reproduced if we introduce new heavy 
isosinglet quarks (and charged leptons) and integrate them out.
These isosinglet quarks can have bare mass terms, and
the new physics scale, $\Lambda_{NEW}$, is identified with that mass.
Such a renormalizable theory is given, in the quark sector, 
by the following Lagrangian terms:
\beq
{\cal L}=-y_{L}\bar{\psi_L}\phi_L U_R-
y_{R}\bar{\psi_R}\phi_R U_L + (h.c.)- M_U {\bar U} U.
\label{eq:4}
\eeq
Using this Lagrangian,
we may compute
the Feynman diagram in which the heavy singlet quark is
exchanged, obtaining for the dimension 5 operator:
\beq
{1 \over \Lambda_{NEW}}= {y_L y_R^{\ast} \over M_U}. 
\label{eq:5}
\eeq 
However this formula leads to the suppression of the 
quark masses by a factor
of  $SU(2)_R$ {\em breaking scale divided by Singlet quark
mass} compared to the electroweak breaking scale, hence it cannot
be applied to the top quark. 
In order to prevent the seesaw mechanism  to act for
the top quark we assume that the corresponding singlet quark has 
a bare mass much smaller than the $SU(2)_R$ breaking
scale. In this case, ignoring flavor mixing for the moment,
the top quark only acquires mass through a dimension 4 Yukawa 
coupling, instead of the mechanism of exchange of a heavy
singlet quark. To illustrate, let us consider the
extreme limit in which the bare mass is set to zero 
in Eq.(\ref{eq:4}), so that we have
\begin{equation}
{\cal L}= - y_L \eta_L \bar{t_L} T_R  - y_R^{\ast} \eta_R
\bar{T_L} t_R + (h.c.).  
\label{eqn:sd}
\end{equation}
These are dimension 4 operators and 
the singlet quark cannot be integrated out because the
bare mass term is absent. The diagonalization is performed 
through maximal mixing for the right-handed sector, as follows: 
\begin{eqnarray}
\left(
\begin{array}{c}
t^m \\ T^m
\end{array}
\right)_L
&=&
\left(
\begin{array}{c}
t \\ T
\end{array}
\right)_L ,   \nn \\
\left(
\begin{array}{c}
t^m \\ T^m
\end{array}
\right)_R
&=&
\left(
\begin{array}{c}
T \\ t
\end{array}
\right)_R,
\end{eqnarray}
(the superscript $m$ denoting the mass states),
leading to: 
\beq
m_t = |y_L| \eta_L, \quad  m_T=  |y_R| \eta_R, 
\label{eq:8}
\eeq
and $m_t$ is of the order of the experimental value.
If we retain the bare mass term, i.e., $ - M_T \bar{T} T $,
 the mass formulae are given by 
\begin{eqnarray}
 m_t &=& |y_L| |y_R| {\eta_L  \eta_R \over \sqrt{|y_R|^2  \eta_R ^2
+{M_T}^2} }, \nn \\
 m_T &=&\sqrt{|y_R|^2 \eta_R ^2 +{M_T}^2}.
\label{eq:8.1}
\end{eqnarray}
\section{Quark Mass Hierarchy and Flavor Mixing}
We will now extend our analysis to the case where flavor mixing
is present and the number of generations is three,
based on the following Lagrangian:
\begin{eqnarray}
{\cal L}&=&-y^{ij}_{0LD}\bar{\psi^{0i}_L}\phi_L
D^j_R-y^{ij}_{0LU}\bar{\psi^{0i}_L}\tilde{\phi_L}U^j_R
-y^{ij}_{0RD}\bar{\psi^{0i}_R}\phi_R
D^j_L-y^{ij}_{0RU}\bar{\psi^{0i}_R}\tilde{\phi_R}U^j_L+ (h.c.)\nonumber\\
&&-y^{ij}_{0LE}\bar{L^{0i}_L}\phi_L
E^j_R-y^{ij}_{0RE}\bar{L^{0i}_R} \phi_R E^j_L+ h.c.\nonumber\\
&&-\bar{U^i}M^i_{\cal U} U^i-\bar{D^i}M^i_{\cal D} D^i-
\bar{E^i}M^i_{\cal E} E^i,\nonumber\\
\label{eq:l0}
\end{eqnarray}
where $M^i_{\cal U}, M^i_{\cal D}$ and $M^i_{\cal E}$
are  real parameters. 
The representations under the gauge groups 
of ordinary quarks and charged leptons
are assigned as:
\bea
\psi_L^{0i}= \left( \begin{array}{c}   u_{0}^{i}\\
                               d_{0}^{i} \end{array} \right)_L
&:& (2,1,1/3),
\quad 
L_L^{0i}= \left( \begin{array}{c}   \nu_{0}^{i}\\
                               e_{0}^{i} \end{array} \right)_L
: (2,1,-1),
\nonumber\\
\psi_R^{0i}=\left( \begin{array}{c}      u_{0}^{i}\\
                              d_{0}^i \end{array} \right)_R
&:&(1,2,1/3),
\quad
L_R^{0i}=\left( \begin{array}{c}      \nu_{0}^{i}\\
                              e_{0}^{i} \end{array} \right)_R
:(1,2,-1),
\nonumber\\
U^i_{L,R}&:&(1,1,4/3),
\quad
D^i_{L,R}:(1,1,-2/3),
\quad
E^i_{L,R}:(1,1,-2).
\label{eq:l1}
\eea
We do not introduce singlet neutrinos and so there are no tree
level neutrino masses.
Let us first study the charged lepton sector and the down quark
sector. The masses of the charged leptons  and down quarks
are much smaller than O($100$GeV), so they can be treated 
in a unified way.
We consider first the down-like quarks.
In order to find the mixing angles (CKM matrix) among both left and right
chiralities, it is convenient to perform
unitary transformations among the ordinary 
quark fields such that the singlet-doublet Yukawa couplings,
 i.e. $y_{0L}$ and $y_{0R}$,  become triangular matrices 
 \cite{MORO}, 
\beq
  U^\dagger y_0 = \left( \begin{array}{ccc} y_{1}   & 0 & 0 \\
                      y_{21} & y_{2} & 0 \\
                      y_{31} & y_{32} & y_{3}
                          \end{array} \right),
\label{eq:l2}
\eeq
where $U$ is a unitary matrix. The diagonal elements in
Eq.(\ref{eq:l2})
are real and the off-diagonal entries are
complex. We perform this same type of transformation in the 
charged lepton sector.
Accordingly, we obtain new bases for the SU(2) doublet quarks and leptons,
\begin{eqnarray}
u_L^\prime &=&U_{UL}^\dagger u_{0L},
\quad
u_R^\prime =U_{RU}^\dagger u_{0R},\nonumber \\
d_L^\prime &=&U_{DL}^\dagger d_{0L},
\quad
d_R^\prime =U_{DR}^\dagger d_{0R},\nonumber \\
e_L^\prime &=&U_{EL}^\dagger e_{0L},
\quad
e_R^\prime=U_{ER}^\dagger e_{0R},\nonumber \\
\nu_L^\prime &=&U_{EL}^\dagger \nu_{0L},
\quad
\nu_R^\prime =U_{ER}^\dagger \nu_{0R}.\nonumber \\
\label{eq:l2.1}
\end{eqnarray}
We note that the transformation on neutrinos is chosen to be
the same as that on charged leptons.  
As a result, the down quark and charged lepton mass
matrices are written as follows:
\begin{equation}
{\cal M_D} =  \left[ \begin{array}
{cccccc}  0  & 0  &0  & y_{LD1} \eta_L & 0 & 0 \\
          0  & 0  &0  & y_{LD21} \eta_L& y_{LD2}\eta_L
  & 0 \\
          0  & 0  &0  & y_{LD31}\eta_L  &
y_{LD32}\eta_L &  y_{LD3} \eta_L\\
         y_{RD1} \eta_R& y_{RD21}^* \eta_R
  &y_{RD31}^*  \eta_R & M_D & 0  &  0\\
          0  & y_{RD2} \eta_R 
& y_{RD32}^* \eta_R
&  0  & M_S  &  0\\
          0  & 0   & y_{RD3} \eta_R
&  0  &  0  &   M_B  \end{array}
    \right],
\label{eq:l2.2}
\end{equation}
\begin{equation}
{\cal M_E} =  \left[ \begin{array}
{cccccc}  0  & 0  &0  & y_{LE1} \eta_L & 0 & 0 \\
          0  & 0  &0  & y_{LE21} \eta_L& y_{LE2}\eta_L
  & 0 \\
          0  & 0  &0  & y_{LE31}\eta_L  &
y_{LE32}\eta_L &  y_{LE3} \eta_L\\
         y_{RE1} \eta_R& y_{RE21}^* \eta_R
  &y_{RE31}^*  \eta_R & M_E & 0  &  0\\
          0  & y_{RE2} \eta_R 
& y_{RE32}^* \eta_R
&  0  & M_{\mu} &  0\\
          0  & 0   & y_{RE3} \eta_R
&  0  &  0  &   M_{\tau}  \end{array}
    \right].
\label{eq:l2.3}
\eeq
These mass matrices can be approximately diagonalized by using the
following
unitary matrices,
\bea
V_{0DL}&=&\left( \begin{array}{cc}  1 & y_{LD} \eta_L
/M_{\cal D} \\
                      -1/M_{\cal D} {y_{LD}}^\dagger \eta_L  & 1 
                         \end{array} \right), \nn \\
V_{0DR}&=&\left( \begin{array}{cc}  1 & y_{RD} \eta_R
/M_{\cal D} \\
                      -1/M_{\cal D} {y_{RD}}^\dagger \eta_R  &1  
                         \end{array} \right),
\label{eq:l2.4}
\eea
where $M=diag(M_D,M_S,M_B)$.
As a result, $\cal M_D$ is approximately diagonalized as
\beq
 V_{0DL}^\dagger {\cal M_D}  V_{0DR} \simeq
 \left( \begin{array}{cccccc} -y_{LD1} \frac{\eta_L \eta_R}{M_D} y_{RD1}
                      &     -y_{LD1}\frac{\eta_L \eta_R}{M_D} y_{RD21}^\ast
                      &     -y_{LD1}\frac{\eta_L \eta_R}{M_D} y_{RD31}^\ast
                      &0&0&0\\
                            -y_{LD21}\frac{\eta_L \eta_R}{M_D} y_{RD1} 
                      &     -y_{LD2}\frac{\eta_L \eta_R}{M_S}y_{RD2} 
                      &     -y_{LD2}\frac{\eta_L \eta_R}{M_S}y_{RD32}^\ast
                      &0&0&0\\
                            -y_{LD31}\frac{\eta_L \eta_R}{M_D} y_{RD1}
                      &     -y_{LD32}\frac{\eta_L \eta_R}{M_S} y_{RD2}
                      &     -y_{LD3}\frac{\eta_L \eta_R}{M_B}y_{RD3}
                      &0&0&0\\
                      0&0&0&M_D&0&0\\
                      0&0&0&0&M_S&0\\
                      0&0&0&0&0&M_B\\
                          \end{array}  \right). 
\label{eq:MD}
\eeq
Approximate eigenvalues of the down quark
masses are given by the diagonal elements of Eq.(\ref{eq:MD}) as:
\begin{eqnarray}
m_b&\cong& \frac{\eta_R}{M_B} y_{LD3} y_{RD3} \eta_L,\nonumber\\
m_s&\cong& \frac{\eta_R}{M_S}
 y_{RD2} y_{LD2} \eta_L,\nonumber\\
m_d&\cong&
\frac{\eta_R}{M_D} y_{RD1} y_{LD1}  \eta_L,\nonumber\\
m_D&\cong& M_D,\nonumber\\
m_S&\cong& M_S,\nonumber\\
m_B&\cong& M_B .
\label{down}
\end{eqnarray}
\noindent
For charged leptons, similar results are obtained.
\subsection{Top quark mass and hierarchy of up type quark masses }
Now we turn to the up type quark masses.
In the previous section we saw that the top quark mass, in the 
absence of flavor mixing, is given by Eq.(\ref{eq:8.1}), rather than
the seesaw type formula.
The extension to the case where flavor mixing is present, and 
$M_T$ is not exactly zero, but restricted to $M_T << \eta_R$, 
was done in \cite{MORO}. Here we extend this previous analysis in
a way that can be applied to the case where $M_T$ is as large as 
$\eta_R$.
First we derive the formulae for the top quark mass
as well as the other light quark masses by solving
the eigenvalue equation.
Consider the up mass matrix:
\begin{equation}
{\cal M_U}=\left( \begin{array}{cc}0&\eta_L y_{0LU}\\
                        \eta_R y_{0RU}^\dagger&M_{\cal U}
           \end{array} 
    \right),
\end{equation}
where $M_{\cal U}=diag(M_U, M_C, M_T)$.
The corresponding eigenvalue equation for the quark masses is given by 
\begin{equation}
\det( {\cal M_U M_U^\dagger}-\Lambda)=\det
\left[ \begin{array}{cc}\eta^2_L y_{0LU} y_{0LU}^\dagger-
\Lambda&y_{0LU} M_{\cal U} \eta_L \\
                        \eta_L M_{\cal U} y_{0LU}^\dagger&\eta_R^2
y_{0RU}^\dagger y_{0RU} + 
M_{\cal U}^2-\Lambda \end{array}
\right]=0.
\label{eig.eq}
\end{equation}
The equation  which determines the eigenvalues  of the order of $
{\eta_L}^2$
(or smaller than  ${\eta_L}^2$)
is reduced to a cubic equation:
\begin{equation}
\det\left[  y_{0LU} \left\{1-M_{\cal U}
(\eta_R^2 {y_{0RU}}^\dagger y_{0RU} + {M_{\cal U}}^2)^{-1} M_{\cal U}
\right\}y_{0LU}^\dagger - \lambda \right]=0,
\label{cub}
\end{equation}
where we use the normalized eigenvalue $ \lambda=
 {\Lambda \over \eta_L^2}$.
We further use the following expansion:
\begin{equation}
(\eta_{0RU}^2 y_{0RU}^\dagger y_{0RU} +M_{\cal U}^2)^{-1}={1\over
M_0}\left[
1-({1\over M_0}\Delta M^2{1\over M_0})+({1\over
M_0}\Delta M^2{1\over M_0})^2+\cdots\right]{1\over M_0},
\label{exp}
\end{equation}
\bea
M_0&=&\left[ \begin{array}{ccc}M_U&\ &\ \nn \\
                              \ &M_C&\ \nn \\
                              \ &\ &
\sqrt{\eta_R^2 (y_{0RU}^\dagger y_{0RU})_{33}+M_T^2}
 \end{array}
     \right]. \\
\label{eq:MO}
\eea
Note that in the traditional treatment of the seesaw model,
Eq.(\ref{exp}) is expanded by the inverse power of the singlet
quark mass matrix, i.e., $diag(M_U,M_C,M_T)$. However the smallness
of $M_T$ compared with $\eta_R$ does not allow us to do so.
 $M_0$ is chosen in such a way that the expansion 
by the inverse power of $M_0$ is still regular even in the
limit of vanishing $M_T$. 
By keeping the dominant coefficients of $\lambda^n
(n=0,1,2,3)$, the eigenvalue equation becomes,
\def\L{{\cal L}}
\def\R{{\cal R}}
\bea
F(\lambda)&=&-\lambda^3 + 
 \{ \L_{33} \R_{33} {X_T}^2 \}  \lambda^2\nn \\
&-& \{ \L_{33} \R_{33}
(\L_{22}-\L_{23}\L_{32}/ \L_{33})
(\R_{22}-\R_{23}\R_{32}/ \R_{33}) {X_T}^2  {X_C}^2 \} \lambda \nn\\
&+&{X_T}^2 {X_C}^2 {X_U}^2 det \L  det \R =0,
\label{eq:eigen} 
\eea
where $\L_{ij} = {(y_{0LU}^\dagger y_{0LU})}_{ij}$,
  $\R_{ij} = {(y_{0RU}^\dagger y_{0RU})}_{ij}$,
$X_U =\frac{\eta_R}{M_U}, X_C =\frac{\eta_R}{M_C},$ and
$ X_T =\frac{\eta_R}{\sqrt{M_T^2 + \R_{33} \eta_R^2}}.$
By noting that $X_U << X_C << X_T =O(1)$, the approximate
solution of Eq.(\ref{eq:eigen}) is
\bea
m_t &= & \sqrt{\L_{33} \R_{33}} {\eta_L \eta_R \over 
\sqrt{{M_T}^2 + \R_{33} \eta_R^2} }, \nn \\
m_c &= &  \sqrt{ (\L_{22}-\L_{23}\L_{32}/ \L_{33})
  (\R_{22}-\R_{23}\R_{32}/ \R_{33})} \{  {\eta_L \eta_R \over
M_C}\}, \nn \\
m_u &= & \sqrt{  \frac {det\L \quad det\R}
 {(\L_{22}-\L_{23}\L_{32}/ \L_{33}) (\R_{22}-\R_{23}\R_{32}/ \R_{33})
       \L_{33} \R_{33} }
} \{ {\eta_L \eta_R \over M_U} \}.  
\label{eqn:up}
\eea
The heavier eigenvalues are also obtained as follows:
\begin{eqnarray}
m_U&=& M_U,\nonumber\\ 
m_C&=& M_C,\nonumber\\
m_T&=& \sqrt{ {M_T}^2 + \R_{33} {\eta_R}^2}.
\label{eqn:hup}
\end{eqnarray}
To compare Eq.(\ref{eqn:up}) and Eq.(\ref{eqn:hup}) with
the formulae obtained in the absence of flavor mixing, 
Eq.(\ref{eq:8.1}), we need to explain a little bit
more.
The mass formulae for the light quarks are obtained in a weak basis in
which
the singlet quark mass matrix $M$ is diagonal, and the 
singlet-doublet Yukawa couplings are general.
However, the formulae are simplified if we go to the
weak basis in which the singlet-doublet Yukawa couplings 
are in the triangular form. For example,
\bea
\sqrt{\L_{33}} &=& y_{LU3}, \quad
\sqrt{\R_{33}} = y_{RU3}, \nn \\
\sqrt{\L_{22}-\L_{23}\L_{32}/ \L_{33}}&=&y_{LU2}, \quad
\sqrt{\R_{22}-\R_{23}\R_{32}/ \R_{33}}=y_{RU2},\nn \\
\sqrt{\det \L}&=& (y_{LU1}) (y_{LU2}) (y_{LU3}),\quad
\sqrt{\det \R}= (y_{RU1}) (y_{RU2}) (y_{RU3}).
\eea
In this basis, the light quark masses are determined by the
diagonal elements of the singlet-doublet Yukawa coupling:
\bea
m_t &= & (y_{LU3}) (y_{RU3}) {\eta_L \eta_R \over 
\sqrt{{M_T}^2 + y_{RU3}^2 \eta_R^2} }, \nn \\
m_c &= &  (y_{LU2}) (y_{RU2}) 
   \{ {\eta_L \eta_R \over M_C} \}, \nn \\   
m_u &= &  (y_{LU1}) (y_{RU1}) 
   \{ {\eta_L \eta_R \over M_U}\} . 
\label{eqn:up2}
\eea
The top quark mass in Eq.(\ref{eqn:up2})
 reduces to the result obtained in Eq.(\ref{eq:8.1}).
Therefore, we conclude that in the triangular basis,
the off-diagonal elements of $y_L$ and $y_R$ can be safely
neglected, since ignoring them does not affect the 
mass eigenvalues significantly. In the following section, it is 
shown that they are related to the
size of the FCNC.
\def\MU{{\cal M_U}}
Finally, we check the formulae Eq.(\ref{eqn:hup})
and Eq.(\ref{eqn:up2}) by
showing that 
$det\MU=m_u m_c m_t m_U m_C m_T $.
In fact, $det \MU$ is given by,
\bea
det {\cal{M_U}} &=& \sum_{p} sign(p)  
\MU_{1P(1)} \MU_{2P(2)} \MU_{3P(3)}
\MU_{4P(4)} \MU_{5P(5)} \MU_{6P(6)}\nn \\
&=&
\MU_{14} \MU_{25} \MU_{36}
\MU_{41} \MU_{52} \MU_{63}\nn \\ &=&
(y_{LU1}) (y_{LU2}) (y_{LU3}) (y_{RU1}) (y_{RU2}) (y_{RU3})
(\eta_L \eta_R)^6.
\label{eq:detM}
\eea
With Eq.(\ref{eqn:hup}), Eq.(\ref{eqn:up2}) and 
$m_T  =  \sqrt{{M_T}^2 + y_{RU3}^2 \eta_R^2}$,
$det \MU $ is equal to the products of mass eigenvalues
of the six quarks.
\def\valpha{{ \mbox{\boldmath$\alpha$}}}
\def\vbeta{{ \mbox{\boldmath$\beta$}}}
\def\vecg{{ \mbox{\boldmath$y$}}}
\def\vecu{{ \mbox{\boldmath$U$}}}
\def\veca{{ \mbox{\boldmath$a$}}}
\def\vecv{{ \mbox{\boldmath$V$}}}
\def\vecu2{{ \mbox{\boldmath$u$}}}
\def\vecv2{{ \mbox{\boldmath$v$}}}   
\def\o{\over}    
\def\Ar{\rightarrow}    
\def\bar{\overline}    
\def\un{\underline}    
\def\r{\gamma}    
\def\d{\delta}    
\def\a{\alpha}    
\def\b{\beta}    
\def\n{\nu}    
\def\m{\mu}    
\def\k{\kappa}    
\def\e{\epsilon}    
\def\p{\pi}    
\def\th{\theta}    
\def\om{\omega}    
\def\vp{{\varphi}}    
\def\Re{{\rm Re}}    
\def\Im{{\rm Im}}    
\def\t{\tilde}    
\def\bar{\overline}    
\def\l{\lambda}    
\def\G{{\rm GeV}}    
\def\Me{{\rm MeV}}    
\def\eV{{\rm eV}}
\def\poa{\psi^{0 \alpha}}
\def\pobal{\bar{\psi^{0 \alpha}}}
\def\poi{\psi^{0 i}}
\def\pobi{\bar {\psi^0 i}}
\def\poLi{{\psi_L}^{0 i}}
\def\poLbi{\bar {{\psi_L}^{0 i}}}
\def\poRi{{\psi_R}^{0 i}}
\def\poRbi{\bar {{\psi_R}^{0 i}}}
\def\guu{\gamma^{\mu}} 
\def\gu{\gamma_{\mu}} 
\def\Au{A^\mu}
\def\ZRu{Z_R^\mu}
\def\ZLu{Z_L^\mu}
\def\sq{\sqrt{g^2+g^{\prime 2}}}
\def\t0{\theta_{20}}
\def\t{\theta_2}
\def\beq{\begin{eqnarray}}
\def\eeq{\end{eqnarray}}
\def\cww{\cos2\theta_W}
\def\cw{\cos\theta_W}
\def\sw{\sin\theta_W}
\def\ssssw{\sin^4\theta_W}
\def\ssw{\sin^2\theta_W}
\def\ccw{\cos^2\theta_W}
\def\ttw{\tan^2\theta_W}
\def\ttww{\tan^4\theta_W}
\def\tw{\tan\theta_W}
\def\nn{\nonumber}
\section{Tree Level Z FCNC}
In this section, we study the tree level FCNC
due to neutral gauge boson exchange. We first 
derive the theoretical form of these flavor changing currents.
This will show how they are naturally suppressed, and
enable us to perform quantitative calculations.
Using the estimates and the present experimental bounds
on rare B and K decays, we examine whether the tree level FCNC
can significantly contribute to them or not.
 We can also compare the tree level FCNC effects with
  one loop GIM suppressed contributions of the standard model.
\\
Let us start with the relevant part of the Lagrangian:
 
\beq
{\cal L}=Z_1 J_1 + Z_2 J_2 + \frac{ {M_1}^2 {Z_1}^2 }{2} +
\frac{ {M_2}^2 {Z_2}^2 }{2}, 
\label{eq:s4.1}
\eeq
where $M_1$ and $M_2$ are the mass eigenvalues of the neutral gauge boson
mass matrix,
\beq
M^2&=&M_{WR}^2   \left( \begin{array}{cc} \frac{\ccw}{\cww}  & 0 \\
                      0 & 0  \end{array} \right) 
     + M_{WL}^2  \left( \begin{array}{cc} \frac{\ssw \ttw}{\cww}  &
                      \frac{\ttw}{{\sqrt{\cww}} } \\
             \frac{\ttw}{{\sqrt{\cww}}\ttw  }& \frac{1}{\ccw}
\end{array}
                      \right), 
\label{eq:s4.2}
\eeq
where $M_{WR}$ and $M_{WL}$ are masses of the 
charged SU(2) gauge bosons and $\sw=e/g$ with 
SU(2) gauge coupling constant g and $U(1)_{em}$
gauge coupling constant e. 
\beq
 \left( \begin{array}{cc} \cos\theta  & \sin\theta \\
                     -\sin\theta & \cos\theta  \end{array} \right)
                   M^2 
\left( \begin{array}{cc} \cos\theta  & -\sin\theta \\
                     \sin\theta & \cos\theta  \end{array} \right)
=\left( \begin{array}{cc} {M_2}^2  & 0 \\
                     0 &  {M_1}^2  \end{array} \right).
\label{eq:s4.3}
\eeq 
When the breaking scale of $SU(2)_R$ is much larger than
that of  $SU(2)_L$, i.e. $M_{WR} >> M_{WL}$, the eigenvalues 
and the mixing angle are approximately given by:
\beq
M_1^2&=&{{M_{WL}}^2 \over \ccw } -
        {{M_{WL}}^4  \over {M_{WR}}^2} 
        {\ttww \over \ccw} , \nn \\
M_2^2&=& {{ {M_{WR}}^2 \ccw} \over {\cww}}, \nn \\
\tan 2\theta &=& {2\sqrt{\cos2\theta_W}  \over \ccw }
 \ttw  { {M_{WL}}^2  \over {M_{WR}}^2 },
\label{eq:s4.4}
\eeq   
and $J_1$ and $J_2$ are written as:
\beq
J_1&=& {J_1}^0 \cos\theta  - {J_2}^0 \sin\theta ,   \nn \\
J_2&=& {J_2}^0 \cos\theta  + {J_1}^0 \sin\theta ,   \nn \\
{J_1}^0&=&\frac {g}{\cw} (-J_{3L} +\ssw J_Q), \nn \\
{J_2}^0&=&g\frac {\cw}{ \sqrt{\cww}} (-J_{3R} -\ttw J_{3L} + \ttw J_Q),
\label{eq:s4.5}
\eeq
where $J_{3L}$, $J_{3R}$, and $J_Q$ are the $SU(2)_L$ isospin
current,  the $SU(2)_R$ isospin current, and the electromagnetic
current, respectively. They are defined as:
\beq
J_{3L \mu}=\sum_{i=1}^3  \poLbi \gu \frac{\tau_3}{2} \poLi, \quad
J_{3R \mu}=\sum_{i=1}^3 \poRbi \gu \frac{\tau_3}{2} \poRi, \quad
J_{Q \mu}=\sum_{\alpha=1}^6 \pobal \gu Q \poa.
\label{eq:s4.6}
\eeq
We write the isospin currents in terms
of the mass eigenstates,
\beq
d_L^{0 i}=\sum_{\alpha=1}^6  V_{DL}^{i \alpha} d_L^{\alpha},
\qquad
d_R^{0 i}=\sum_{\alpha=1}^6  V_{DR}^{i \alpha} d_R^{\alpha},
\label{eq:s4.7}
\eeq
where $V_{DL}$ and $V_{DR}$ are 
the $ 6 \times 6$ unitary matrices 
that diagonalize the 
seesaw type mass matrix,
$$
{V_{DL}}^\dagger \left( \begin{array}{cc} 0  & y_{0LD} \eta_L \\
                      {y_{0RD}}^\dagger \eta_R & M_{\cal D}
                         \end{array}  \right) V_{RD}
=\left( \begin{array}{cc}  m_d & 0 \\
                      0 &  m_D  \end{array} \right).
\label{eq:s4.8}
$$
Now the neutral currents in the down quark sector are written as:
\beq
{J_{3L}}&=& - \frac{1}{2}
\sum_{\alpha  \beta} {\cal Z}_{DL \alpha
\beta} {\bar{{d_L}^\alpha}}\gu {d_L}^\beta, \nn \\
{J_{3R}}&=& - \frac{1}{2}  
\sum_{\alpha  \beta} {\cal Z}_{DR \alpha
\beta} {\bar{{d_R}^\alpha}}\gu {d_R}^\beta,
\label{eq:s4.9}
\eeq
where
\beq
{\cal Z}_{DL  \alpha \beta}&=& 
\sum_{i=1}^{3} V_{DL}^{\dagger \alpha i} V_{DL}^{i \beta}
=\delta^{\alpha \beta}-\sum_{I=4}^{6}  
V_{DL}^{\dagger \alpha I} V_{DL}^{I \beta}, \nn \\
{\cal Z}_{DR \alpha \beta}&=&
\sum_{i=1}^{3} V_{DR}^{\dagger \alpha i} V_{DR}^{i \beta}
=\delta^{\alpha \beta}- \sum_{I=4}^{6} 
V_{DR}^{\dagger \alpha I} V_{DR}^{I \beta}.
\label{eq:s4.10}
\eeq
In this way the tree level exchange of neutral gauge bosons
gives rise to the effective $\Delta F=1$
Lagrangian,
\beq 
{\cal L}^{\Delta F=1}&=&\sqrt{2} G_F  \sum_{\alpha \ne \beta}
\{ {\cal Z}_{DL \alpha \beta}
\bar{{d_L}^\alpha} \gu {d_L}^\beta \bar{{\nu_L}^i} \guu
{\nu_L}^i  
+ \beta {\cal Z}_{DR \alpha \beta}
\bar{{d_R}^\alpha} \gu {d_R}^\beta \bar{{\nu_R}^i} \guu
{\nu_R}^i \} \nn  \\ 
&-&\sqrt{2} G_F \sum_{\alpha \ne \beta} \{ {\cal Z}_{DL \alpha
\beta} \bar{{d_L}^\alpha} \gu {d_L}^\beta (\bar{{l_L}^i}
\guu {l_L}^i - 2 \ssw \bar{{l}^i} \guu {l}^i) \nn 
\\ 
&+&\beta {\cal Z}_{DR \alpha \beta} \bar{{d_R}^\alpha} \gu
{d_R}^\beta (\bar{{l_R}^i} \guu {l_R}^i - 2 \ssw \bar{{l}^i}
\guu {l}^i) \},
\label{eq:s4.11}
\eeq
where the coefficient
 $\beta$ is defined as $\beta = M_{W_L}^2/M_{W_R}^2$.  As we
discuss below, the strength of the tree level FCNC couplings 
${\cal Z}_{DR}$ are enhanced by a factor of $1/\beta$ compared
to ${\cal Z}_{DL} $.  Therefore the $\Delta F=1$ FCNC due to 
$Z_1$ is of the same order of magnitude as that of $Z_2$.
\\ 
\def\BXsee{B \rightarrow X_s e^+ e^-}
\def\BXsnunu{B \rightarrow X_s  \nu \bar{\nu}}
\def\BXdee{B \rightarrow X_d e^+ e^-}
\def\BXnunu{B \rightarrow X  \nu \bar{\nu}}
\def\BXdnunu{B \rightarrow X_d  \nu \bar{\nu}}
\def\Kpienu{K^+ \rightarrow \pi^0 e^+ {\nu}}
\def\Kpiee{K^+ \rightarrow \pi^+ e^+ e^-}
\def\Kpinunu{K^+ \rightarrow \pi^+ \nu \bar{\nu}}
\def\KLpiee{K_L \rightarrow \pi^0 e^+ e^-}
\def\KLpinunu{K_L \rightarrow \pi^0 \nu \bar{\nu}}
\def\BXclnu{B \rightarrow X_c l^- \bar{\nu}}
We now turn to theoretical estimate.
We first show how the FCNC 
among the ordinary down type quarks is suppressed.
It turns out that the suppression factor is $ {m_{d}}^i
{m_{d}}^j/ {M_{W_R}}^2 $
for ${\cal Z}_{DL}$ and  $ {m_{d}}^i {m_{d}}^j
/{M_{W_L}}^2 $ for ${\cal Z}_{DR}$.
As seen in Eq.(\ref{eq:s4.11}), 
the FCNC couplings among the ordinary
quarks are given by:
\begin{eqnarray}
{\cal Z}_{DLij}&=&-V_{DL}^{\dagger iI}V_{DL}^{Ij},\nn 
\\
{\cal Z}_{DRij}&=&-V_{DR}^{\dagger iI}V_{DR}^{Ij}.
\label{eq:s4.12}
\end{eqnarray}
In order to estimate the FCNC using Eq.(\ref{eq:s4.12}),
we must know $V_{DL}$ and $V_{DR}$.
For this purpose we can
start with an approximate parameterization obtained
from the diagonalization of the seesaw mass matrix and unitarity
(see Eq.(\ref{eq:l2}) and Eq.(\ref{eq:l2.1})), then we get
\bea
V_{DL}&=&\left(  \begin{array}{cc}  U_{DL} & 0 \\
                               0  & 1 \end{array} \right)
V_{0DL} 
=\left( \begin{array}{cc}  U_{DL}  & y_{0LD} \eta_L /M_{\cal
D}\\
                      -1/M_{\cal D} {y_{0LD}}^\dagger 
                      \eta_L U_{DL} & 1 
                         \end{array} \right), \nn \\
V_{DR}&=&\left(  \begin{array}{cc}  U_{DR} & 0 \\
                               0  & 1 \end{array} \right)
V_{0DR}
=\left( \begin{array}{cc} U_{DR}  & y_{0RD} \eta_R /M_{\cal D} \\
                      -1/M_{\cal D} {y_{0RD}}^\dagger \eta_R U_{DR} &1  
                         \end{array} \right),
\label{eq:s4.13}
\eea
where $U_{DL}$ and $U_{DR}$ are unitary matrices that approximately 
diagonalize the $3 \times 3$ effective light quark mass matrix:
\beq
-U_{DL}^\dagger \left( y_{0LD} 
{\eta_L \eta_R \over M_{\cal D}} y_{0RD}^\dagger \right) U_{DR} = {m_d}.
\label{eq:s4.14}
\eeq
Notice that with a suitable choice of
the unitary matrix $U$,  an arbitrary matrix $y_0$  
can be transformed into a triangular matrix, as in Eq.(\ref{eq:l2}).
Suppose   $U_{DL}^\dagger y_{0DL}$ and  $U_R^\dagger y_{0DR} $
are such triangular matrices.
We immediately find that the effective light quark
mass matrix is approximately diagonalized,
\beq
-U_{DL}^\dagger (y_{0LD} {\eta_L \eta_R \over M_{\cal D}}
y_{0RD}^\dagger) U_{DR} =
-  \left( \begin{array}{ccc} y_{DL1}\frac{1}{M_D} y_{DR1}   
                      &     y_{DL1}\frac{1}{M_D} y_{DR21}^\ast
                      &     y_{DL1}\frac{1}{M_D} y_{DR31}^\ast \\
                              y_{DL21}\frac{1}{M_D} y_{DR1}  
                      &     y_{DL2}\frac{1}{M_S} y_{DR2} 
                      &     y_{DL2}\frac{1}{M_S} y_{DR32}^\ast \\
                            y_{DL31}\frac{1}{M_D} y_{DR1}  
                      &     y_{DL32}\frac{1}{M_S} y_{DR2} 
                      &     y_{DL3}\frac{1}{M_B} y_{DR3}  
                          \end{array}  \right) \eta_L \eta_R,
\label{eq:s4.15}
\eeq
where we assume the following hierarchy of the singlet quark
masses,
\beq
 M_D \gg M_S \gg M_B.
\label{eq:s4.16}
\eeq
 Therefore the light quark
masses are approximately given by the diagonal elements of
Eq.(\ref{eq:s4.15}),
\beq
m_d^i=-\left( y_{DLi} \frac{1}{M_{{\cal D} i}}  y_{DRi} \right)
\eta_L \eta_R.
\label{eq:s4.17}
\eeq
 Substituting the approximate parameterization of $V_{DL}$ and 
$V_{DR}$ into Eq.(\ref{eq:s4.13}), we obtain the
 following formulae for the FCNC:
\begin{eqnarray}
{\cal Z}_{DL}^{ij}&=&-\{ U_{DL}^\dagger y_{0LD} 
{\eta_L^2 \over
{M_{\cal D}}^2} y_{0LD}^\dagger
U_{DL} \}_{ij} \nonumber \\
          &=&-{m_d}^i  { \{ U_{DR}^\dagger  y_{0RD}
y_{0RD}^\dagger 
U_{DR}\}^{-1}}_{ij}
 {m_d}^j/ {\eta_R}^2,
          \nonumber \\
{\cal Z}_{DR}^{ij}&=& -{m_d}^i { \{ U_{DL}^\dagger y_{0LD} 
y_{0LD}^\dagger U_{DL}\}^{-1}}_{ij}
{m_d}^j/ {\eta_L}^2.
\label{eq:s4.18}
\end{eqnarray}
We must check that the right hand side does not necessarily vanish when 
 $i \ne j$.
We can see that this is indeed the case because  $U_L^\dagger
y_{0L}$
and $U_R^\dagger y_{0R}$
are triangular matrices, and their off diagonal
elements, $y_{ij L}$ and $y_{ij R}$ $( i \ne  j)$, are
non-zero in general.
To leading order the off diagonal
elements are given by:
\bea
 \{ U_{DR}^\dagger y_{0RD} y_{0RD}^\dagger U_{DR} \}^{-1}_{ij}
&\simeq& \Bigl\{\ \begin{array}{c}  
-{y_{RDji}^\ast \over y_{RDi} y_{RDj}^2} \quad (i
< j),  \\
 -{y_{RDij} \over y_{RDj} y_{RDi}^2} \quad (i
> j), \end{array} \nn \\
 \{U_{DL}^\dagger y_{0LD} y_{0LD}^\dagger U_{DL} \}^{-1}_{ij}
 &\simeq& \Bigl\{ \begin{array}{c}
-{y_{DLji}^\ast \over y_{LDi} y_{LDj}^2} \quad (i
< j),\\
-{y_{DLij} \over y_{LDj} y_{LD
i}^2} \quad (i
> j),\end{array}
\eea
so the FCNC couplings are suppressed as:
\begin{eqnarray}
{\cal Z}_{DL ij}&=&- \{ {m_d}^i {m_d}^j /{2 M_{W_R}}^2 \} g^2
 (U_{DR}^\dagger  y_{RD} y_{RD}^\dagger U_{DR})^{-1}_{ij} \nonumber \\
 &\simeq& ({m_d}^i {m_d}^j /{2 M_{W_R}}^2) 
\frac{y_{RDji}\ast}{y_{RDi}} \left(\frac{g}{y_{RDj}}\right)^2
\quad (i < j),\nn \\
{\cal Z}_{DL ij}&=&{\cal Z}_{DL ji}^{\ast}  \quad (i > j), \nn \\
{\cal Z}_{DR ij}&=&-\{ {m_d}^i {m_d}^j /{2 M_{W_L}}^2 \} g^2
 (U_{DL}^\dagger  y_{LD} y_{LD}^\dagger U_{DL})^{-1}_{ij} \nn \\
 &\simeq& ({m_d}^i {m_d}^j /{2 M_{W_L}}^2)
\frac{y_{LDji}\ast}{y_{LDi}} \left(\frac{g}{y_{LDj}}\right)^2
    \quad ( i < j), \nn \\
{\cal Z}_{DR ij}&=&{\cal Z}_{DR ji}^{\ast}  \quad (i > j).
\label{eq:s4.19}
\end{eqnarray}
Therefore the tree level FCNC among the light quarks 
is naturally suppressed by a factor of ({\em quark masses})$^2$
{\em divided by an} $(SU(2)$ {\em breaking scale}$)^2$. 
 Also, they are proportional to the
off-diagonal elements of $y_{ij}$. If all of these vanish, 
there are
no FCNC among light quarks.\\
We can now confront our result for the FCNC with the present 
experimental bounds from rare K and B decays.
\beq
{\cal Z}_{R sd}&\simeq& {m_s m_d \over  2 {M_{WL}}^2}
\frac{y_{Lsd}}{y_{Ld}} \left(\frac{g}{y_{Ls}}\right)^2= 3.1
\times 10^{-7} \{ \frac{y_{Lsd}}{2 y_{Ld}}
 \left(\frac{g}{y_{Ls}}\right)^2 \},     \nn \\
{\cal Z}_{R bd}&\simeq& {m_b m_d \over  2 {M_{WL}}^2}
 \frac{y_{Lbd}}{y_{Ld}} \left(\frac{g}{y_{Lb}}\right)^2 
= 7.8\times 10^{-6} \{
\frac{y_{Lbd}}{2 y_{Ld}} \left(\frac{g}{y_{Lb}}\right)^2
\},
\nn \\
{\cal Z}_{R bs} &\simeq& {m_b m_s \over 2  {M_{WL}}^2}
\frac{y_{Lbs}}{y_{Ls}} \left(\frac{g}{y_{Lb}}\right)^2
= 1.6\times 10^{-4}  
\{ \frac{y_{Lbs}}{2 y_{Ls}}
\left(\frac{g}{y_{Lb}}\right)^2 \},          \nn \\
{\cal Z}_{L sd}&\simeq&  \beta {m_s m_d \over  2 {M_{WR}}^2}
\frac{y_{Rsd}}{y_{Rd}} \left(\frac{g}{y_{Rs}}\right)^2= 1.2
\times 10^{-8} \left(\frac{400} {M_{WR}(GeV)} \right)^2 
\{ \frac{y_{Rsd}} {2 y_{Rd}} \left(\frac{g}{y_{Rs}}\right)^2
\}, \nn \\
{\cal Z}_{L bd}&\simeq& \beta {m_b m_d \over  2 {M_{WR}}^2}
\frac{y_{Rbd}}{y_{Rd}} \left(\frac{g}{y_{Rb}}\right)^2
= 3.1 \times 10^{-7} \left(\frac{400} {M_{WR}(GeV)}
\right)^2
\{ \frac{y_{Rbd}} {2 y_{Rd}} \left(\frac{g}{y_{Rb}}\right)^2
\}, \nn \\
{\cal Z}_{L bs} &\simeq& \beta {m_b m_s \over 2  {M_{WR}}^2}
\frac{y_{Rbs}}{y_{Rs}} \left(\frac{g}{y_{Rb}}\right)^2
= 6.4 \times 10^{-6} \left(\frac{400} {M_{WR}(GeV)} \right)^2 
\{ \frac{y_{Rbs}} {2 y_{Rs}} \left(\frac{g}{y_{Rb}}\right)^2
\},\nn \\
\beta &=&{M_{WL}^2 \over M_{WR}^2},
\label{eq:s4.20}
\eeq
where we use $m_d=10$ MeV,  $m_s=0.2$ GeV, $m_b=5$ GeV
and
$M_{W_L}=80$ GeV.
We also use the following notations.
${\cal Z}_{Lbs} \equiv {\cal Z}_{DL32}, y_{Rbs} \equiv 
y_{RD32}$ and so
on. 
The tree level contribution to the branching ratio 
$ \BXsee$ and  $\Kpinunu$ is given by, 
\bea
{ {\cal B}(\Kpinunu)|_{\rm Tree} \over {\cal B} (\Kpienu)}&=&
\frac{3}{2} { |{\cal Z}_{Lsd}|^2+\beta^2  |{\cal Z}_{Rsd}|^2  \over
|V_{us}|^2    },
\label{eq:s4.22}\\
{ {\cal B}(\BXsee)|_{Tree} \over {\cal B} (\BXclnu)} &=&{P_s \over
P_c}\frac{1}{4} \{ (1-2\ssw)^2+ 4 \ssssw\} {  |{\cal
Z}_{Lsb}|^2+\beta^2  |{\cal Z}_{Rsb}|^2  \over
|V_{cb}|^2    },
\label{eq:s4.21}
\eea
where $P_c = 0.538 $ and $P_s = 0.986$ are 
the phase space factors for the charm quark and the strange quark, 
respectively.
Using ${\cal B} (\BXclnu) \simeq 10 \%$, $|V^{\rm CKM}_{cb}|=0.04$,
${\cal B}(\Kpienu) \simeq 5 \% $ and $|V^{\rm CKM}_{us}|=0.22$ \cite{PDG},
 the order of the magnitude of the tree level contribution
to these processes are,
\bea
 {\cal B}(\BXsee)|_{\rm Tree}&\le& 1.4 \times 10^{-9}, \nn \\ 
 {\cal B}(\Kpinunu)|_{\rm Tree}&\le& 4.5 \times 10^{-16}, 
\label{eq:s4.23}
\eea
where $M_{WR} \geq 400 \rm(GeV)$ and the combination
of the coefficient of Yukawa coupling and gauge coupling 
in Eq.(\ref{eq:s4.20}) 
is set to be ${\rm O}(1)$. The experimental bound \cite{CLEO}  
and  the prediction of the standard model \cite{AHHM} of $\BXsee$
are,
\bea
{\cal B}(\BXsee)|_{\rm Exp.}  &\le & 5.7 \times 10^{-5}, \nn \\ 
{\cal B}(\BXsee)|_{\rm SM} &\simeq&  8.4 \pm 2.3  \times   10^{-6},  
\label{eq:s4.24}
\eea
while the recent measurement of $\Kpinunu$  \cite{E787} and the
prediction of the standard model \cite{BURA} are given by,
\bea
{\cal B}(\Kpinunu)|_{\rm Exp.}&=&4.2^{+9.7}_{-3.5}\times
10^{-10},\nn
\\  
{\cal B}(\Kpinunu)|_{\rm SM} &=& 0.6 \sim 1.5 \times 10^{-10}.
\label{eq:s4.25}
\eea
Compared with  values in Eqs.(\ref{eq:s4.23}),
(\ref{eq:s4.24}) and 
(\ref{eq:s4.25}),
  the tree level FCNC is found to be negligibly small.
In section 6 and 7, we investigate the FCNC in one-loop level.
\def\valpha{{ \mbox{\boldmath$\alpha$}}}
\def\vbeta{{ \mbox{\boldmath$\beta$}}}
\def\vecg{{ \mbox{\boldmath$y$}}}
\def\vecu{{ \mbox{\boldmath$U$}}}
\def\veca{{ \mbox{\boldmath$a$}}}
\def\vecv{{ \mbox{\boldmath$V$}}}
\def\vecu2{{ \mbox{\boldmath$u$}}}
\def\vecv2{{ \mbox{\boldmath$v$}}}   
\def\o{\over}    
\def\Ar{\rightarrow}    
\def\bar{\overline}    
\def\un{\underline}    
\def\r{\gamma}    
\def\d{\delta}    
\def\a{\alpha}    
\def\b{\beta}    
\def\n{\nu}    
\def\m{\mu}    
\def\k{\kappa}    
\def\e{\epsilon}    
\def\p{\pi}    
\def\th{\theta}   
\def\om{\omega}    
\def\vp{{\varphi}}    
\def\Re{{\rm Re}}    
\def\Im{{\rm Im}}    
\def\t{\tilde}    
\def\bar{\overline}    
\def\l{\lambda}    
\def\G{{\rm GeV}}    
\def\Me{{\rm MeV}}    
\def\eV{{\rm eV}} 
\def\yuUp{y11'}
\def\yuCp{y12'}
\def\yuTp{y13'}
\def\ycTp{y23'}
\def\ycUp{y11'}
\def\ycCp{y22'}
\def\ytUp{y31'}
\def\ytCp{y32'}
\def\ytTp{y33'}
\def\U{{\cal U}}   
\section{The flavor mixing in the charged currents}
In this section, we present the approximate parameterization
of the flavor
mixing in the charged currents
which is analogous to CKM
matrix in the standard model (SM).
The difference between the CKM in the standard
model and that of the present model is
 that the flavor mixing  in the present
model is $ 6 \times 6 $ rather than $ 3 \times 3$
in the SM. There is also right-handed flavor
mixing in the
present model.
As we show in the following,
the flavor mixing consists of 
the singlet-doublet mixing and
the flavor mixing among doublet quarks.
The charged currents of the present model are  written as,
\beq
{J_{L\mu}}&=&{\bar{{u_L}^\alpha}}\gu V^L_{\alpha \beta} {d_L}^\beta, \nn \\
{J_{R\mu}}&=&{\bar{{u_R}^\alpha}}\gu V^R_{\alpha \beta}
{d_R}^\beta, 
\eea 
where $V^L$ and $V^R$ are the generalized CKM matrix and are given by
\bea
V^L_{\alpha \beta}&=&\sum_i^3 V^\dagger_{UL\alpha i}V_{DL i
\beta}
=\sum_i^3  V^\dagger_{0UL\alpha i}
U^L_{ij}  V_{0DL j \beta},    \nn \\
V^R_{\alpha \beta}&=&\sum_i^3 V^\dagger_{UR\alpha i}
V_{DR i \beta}=\sum_i^3  V^\dagger_{0UR\alpha i}
U^R_{ij}  V_{0DR j \beta},
\eea
where   $U^L_{ij}$ and $U^R_{ij}$ are 3 by 3 unitary
matrices and are  defined by,
\bea
U^L &=&U_{UL}^\dagger U_{DL}, \nn \\
U^R &=&U_{UR}^\dagger U_{DR}.
\eea
$U_{UL}, U_{UR}, U_{DL},$ and $U_{DR}$ are 
$3 \times 3$ unitary matrices which transform 
the singlet-doublet Yukawa matrices 
$ y_{0UL}, y_{0UR}, y_{0DL},$ and $y_{0DR}$
into the triangular matrices,
$ y_{UL}, y_{UR}, y_{DL},$ and $y_{DR}$.\\
 $V_{0UL/R} (V_{0DL/R})$  are matrices which diagonalize $
{\cal M_U}$
and    ${\cal M_D}$ as,
\bea
 V_{0UL}^\dagger {\cal M_U}  V_{0UR}
 =\left( \begin{array}{cc}  m_u & 0 \\
                      0 &  m_U  \end{array} \right), \nn \\ 
 V_{0DL}^\dagger {\cal M_D}  V_{0DR} =\left( \begin{array}{cc}  m_d & 0 \\
                      0 &  m_D  \end{array} \right),
\eea
where  
\bea
{\cal M_U}= \left( \begin{array}{cc} 0 & y_{UL} \eta_L \\
                               y_{UR}^\dagger 
                               \eta_R & M_{\cal U}
\end{array} \right) , \nn \\
{\cal M_D}= \left( \begin{array}{cc} 0 & y_{DL} \eta_L \\
                               y_{DR}^\dagger 
                               \eta_R & M_{\cal D}
\end{array} \right). 
\eea
Because the mass eigenvalues are not affected significantly
by the
presence of the off-diagonal element of the triangular 
matrices $y_{UL}$,$y_{UR}$, $y_{DL}$ and $y_{DR}$,
it is legitimate to neglect
the off-diagonal element of the triangular 
matrices. In the approximation,  ${\cal M_D}$ and 
 ${\cal M_U}$ are
approximated as,
\bea
 {\cal M_U}  &\simeq& \left[ \begin{array}
{cccccc}  0  & 0  &0  & y_{LU1} \eta_L & 0 & 0 \\
          0  & 0  &0  & 0&  y_{LU2}  \eta_L & 0 \\
          0  & 0  &0  & 0  & 0 & y_{LU3} \eta_L\\
         y_{RU1}  \eta_R& 0  & 0& M_U & 0  &  0\\
          0  & y_{RU2} \eta_R &  0 &  0  & M_C  &  0\\
          0  & 0   & y_{RU3} \eta_R&  0  &  0  &   M_T
  \end{array}     \right], \nn \\
 {\cal M_D}  &\simeq& \left[ \begin{array}
{cccccc}  0  & 0  &0  & y_{LD1} \eta_L & 0 & 0 \\
          0  & 0  &0  & 0&  y_{LD2}  \eta_L & 0 \\
          0  & 0  &0  & 0  & 0 & y_{LD3} \eta_L\\
         y_{RD1}  \eta_R& 0  & 0& M_D & 0  &  0\\
          0  & y_{RD2} \eta_R &  0 &  0  & M_S  &  0\\
          0  & 0   & y_{RD3} \eta_R&  0  &  0  &   M_B
  \end{array}     \right].
\label{matrix}
\eea
It may be useful to comment on the diagonal forms of the matrices in 
Eq.(\ref{matrix}).
These simple block-diagonal forms follow from
 neglecting the off diagonal entries
 of the initial $6\times 6$ quark mass matrices as in Eq.(\ref{matrix}).
Therefore, the forms of Eq.(\ref{matrix}) become more complex in general if
 the off diagonal entries  of the triangular matrices
 denoted by $y_L$ and $y_R$  
 of the quark mass matrices are included. 
 Because the tree level FCNC is already suppressed,
 in one loop level calculation, we can set
 $ Z_{ij}=0$. This is equivalent to neglecting
 the off-diagonal element of the triangular matrices
$y_{ij}$ in the one-loop calculation.
 Therefore, it is sufficient to keep only the
diagonal element of $y$ for the present purpose.
In the approximation, there is a convenient parameterization
of the $ 6\times 6$ generalized KM.
First the matrices shown in Eq.(\ref{matrix})
 can be diagonalized by the 2 by 2 block
diagonal matrices, because we only need to diagonalize 
a singlet-doublet mixing in each flavor, 
\bea
V_{0LU} &=& \left( \begin{array}{cc} C_{LU} & S_{LU}  \\
                                  -S_{LU} &  C_{LU}
\end{array} \right), \quad 
V_{0RU} = \left( \begin{array}{cc} C_{RU} & S_{RU}  \\
                                  -S_{RU} &  C_{RU}
\end{array} \right),\nn \\
V_{0LD} &=& \left( \begin{array}{cc} C_{LD} & S_{LD}  \\

                                  -S_{LD} &  C_{LD}
\end{array} \right),
\quad
V_{0RD} = \left( \begin{array}{cc} C_{RD} & S_{RD} \\
                                  -S_{RD} &  C_{RD} 
                                   \end{array}\right),
\eea 
where,
\bea
 S_{LU} &=& 
 \left 
(\matrix{s_{Lu} & 0 & 0 \cr 0 & s_{Lc} & 0 \cr 0 & 0 &
s_{Lt}\cr}\right ),\quad
 C_{LU}  = 
 \left (
\matrix{c_{Lu} & 0 & 0 \cr 0 & c_{Lc} & 0 \cr 0 & 0 &
c_{Lt}\cr}\right ), \nn \\
 S_{RU} &=& 
 \left 
(\matrix{s_{Ru} & 0 & 0 \cr 0 & s_{Rc} & 0 \cr 0 & 0 &
s_{Rt}\cr}\right ),\quad
 C_{RU} = 
 \left (
\matrix{c_{Ru} & 0 & 0 \cr 0 & c_{Rc} & 0 \cr 0 & 0 &
c_{Rt}\cr}\right ), \nn \\
 S_{LD} &=& 
 \left 
(\matrix{s_{Ld} & 0 & 0 \cr 0 & s_{Ls} & 0 \cr 0 & 0 &
s_{Lb}\cr}\right ),\quad
 C_{LD}  = 
 \left (
\matrix{c_{Ld} & 0 & 0 \cr 0 & c_{Ls} & 0 \cr 0 & 0 &
c_{Lb}\cr}\right ), \nn \\
S_{RD} &=& 
 \left 
(\matrix{s_{Rd} & 0 & 0 \cr 0 & s_{Rs} & 0 \cr 0 & 0 &
s_{Rb}\cr}\right ),\quad
 C_{RD} = 
 \left 
(\matrix{c_{Rd} & 0 & 0 \cr 0 & c_{Rs} & 0 \cr 0 & 0 &
c_{Rb}\cr}\right ),
\eea
\label{Qqmix}
with $s_{Lu}=\sin\theta_{Lu}$ and $c_{Lu}=\cos\theta_{Lu}$ etc.
and likewise for the down-quark sector.
The flavor mixings of left handed quarks are given by the following 
$6\times 6$ matrix:
\begin{equation}
V^L = \left (\matrix{C_{LU}  & -S_{LU} \cr
  S_{LU}  & C_{LU}\cr }\right ) \left (\matrix{ U^L  & 0 \cr  0 & 0 \cr
}\right )
\left (\matrix{C_{LD}  & S_{LD} \cr  -S_{LD}  & C_{LD}\cr }\right ) = 
\left (\matrix{C_{LU} U^L C_{LD} & C_{LU} U^L S_{LD}\cr
  S_{LU} U^L C_{LD} & S_{LU} U^L S_{LD}\cr }\right ) \ , 
\end{equation}
where
 $3\times 3$ part of the generalized
 KM matrix $V^L$ is no longer 
unitary, while $U^L$ is an unitary matrix.   
For the right-handed quarks, the CKM matrix is given as follows:
\begin{equation}
V^R=
 \quad\left (\matrix{C_{RU} U^R C_{RD} & C_{RU} U^R S_{RD}\cr
  S_{RU} U^R C_{RD} & S_{RU} U^R S_{RD}\cr  }\right ) \ .
\end{equation}
We  give an approximate parameterization of $V^L$ and $V^R$.
We first note that the mixings $s_{Li}$ and $s_{Ri}$
 are  expressed in terms of the mass matrices elements:
\begin{equation}
 s_{Lu} \simeq \frac{y_{LU1}\eta_L}{M_U}  , \quad
 s_{Lc} \simeq \frac{y_{LU2}\eta_L}{M_C}, \quad
 s_{Lt} =\sin \left [\frac{1}{2}\tan^{-1} \frac{2|y_{LU3}|\eta_L
M_T}{|y_{RU3}|^2\eta_R^2+ 
        M_T^2-|y_{LU3}|^2\eta_L^2}\right ] ,
\label{eq:sL}
\end{equation}
\begin{equation}
 s_{Ru} \simeq \frac{y_{RU1}\eta_R}{M_U}  , \quad
 s_{Rc} \simeq \frac{y_{RU2}\eta_R}{M_C}, \quad
 c_{Rt} =\sin \left [\frac{1}{2}\tan^{-1}\frac{2|y_{RU3}|\eta_R
M_T}{|y_{RU3}|^2\eta_R^2- 
  M_T^2-|y_{LU3}|^2\eta_L^2} \right ] .
\end{equation}
There is a simple relation among the mixing angles,
the light quark mass $m_{qi}$ and heavy quark mass 
$m_{Qi}$ which follows from zeros of the 
$3\times3$ ordinary light quark sector of 
the mass matrix,
\bea
c_{Ri}c_{Li} m_{qi}=-s_{Ri}s_{Li} m_{Qi}.
\label{relation}
\eea
These relations are useful to estimate the box contributions
in FCNC.
We observe the following facts on the mixing angles
between the singlet and doublet quarks.
\begin{itemize}
\item{ The mixing angles are 
suppressed as   $s_{Li} =O(m_i/M_{WR})$
and $s_{Ri} =O(m_i/M_{WL})$ for 
light five quarks (from up to bottom quarks )and 
their singlet partners.}
\item{
$s_{Lt} $ is
 at the most $O({ M_{WL} \over M_{WR}})$ 
 with $M_T=O(\eta_R)$.
For  $M_{WR} > 400 (GeV)$, $s_{Lt}$ may
be less than $0.2$.}
\item{ $s_{Rt}$ is not suppressed. In section 2,
we argue that when $M_T < M_{WR}$, we can prevent the 
seesaw mechanism to act and the top quark mass
is at the electroweak breaking scale without introducing
non-perturbative Yukawa coupling. This corresponds to 
$c_{Rt}=O({M_T \over M_{WR}}) <1 $ and $s_{Rt}$ is close to 
one. }
\end{itemize} 
Because $U^L$ is $ 3\times 3 $ unitary matrix
and we have direct measurements on $V^L_{ui}$ and
$V^L_{ci}$, we can parameterize $U^L$ with the Wolfenstein
parameterization. 
About the right-handed KM, we are guided by theoretical
assumption. We assume that the Yukawa coupling between
the doublet and the singlet quarks is left-right symmetric,
$y_L=y_R$. This reduces to $U^L=U^R$.
In the following section, we always assume this relation.
It is instructive to write the generalized KM $V^L$ and $V^R$
explicitly in the approximate parameterization.
We are interested in the charged currents between 
light down type quarks (d,s,b) and up type quarks.
By neglecting the small mixings to the heavy quarks (U,D,S,B,C)
, 
$ 4 \times 3 $ part of the $V^L$ and $V^R$ are:
\bea
\left( \begin{array}{cccc}
{\bar u_L} {\bar c_L} {\bar t_L} {\bar T_L} \end{array}
\right)     
 \left( \begin{array}{ccc}  
1-\frac{\lambda^2}{2} &
\lambda & A \lambda^3 (\rho-i \eta)   \\
-\lambda-i A^2 \lambda^5 \eta & 1- \frac{\lambda^2}{2} & A \lambda^2  \\
c_{Lt} A \lambda^3 (1-\rho-i \eta) &  c_{Lt} (-A-iA \lambda^4 \eta) 
\lambda^2 & c_{Lt}  \\
s_{Lt} A \lambda^3 (1-\rho-i \eta) &  s_{Lt} (-A-i A \lambda^4 \eta) 
\lambda^2 & s_{Lt} \end{array}\right) 
 \left(\begin{array}{c} d_L  \\ s_L  \\ b_L 
 \end{array}
 \right),
\eea
\bea
\left(\begin{array}{cccc}{\bar u_R} {\bar c_R} {\bar t_R}
 {\bar T_R} \end{array} \right)     
 \left( \begin{array}{ccc}  
1-\frac{\lambda^2}{2} &
\lambda & A \lambda^3 (\rho-i \eta)   \\
-\lambda-iA^2 \lambda^5 \eta  & 1- \frac{\lambda^2}{2} & A \lambda^2 \\
c_{Rt} A \lambda^3 (1-\rho-i \eta) &  c_{Rt} (-A 
\lambda^2-i A \lambda^4 \eta) & c_{Rt}  \\ 
s_{Rt} A \lambda^3 (1-\rho-i \eta) & s_{Rt} (-A 
\lambda^2-i A \lambda^4 \eta)  & s_{Rt} \end{array} \right) 
 \left(\begin{array}{c} d_R  \\ s_R  \\ b_R 
 \end{array}
 \right). 
 \label{eq:KM}
\eea
Note that $V^L$ and $V^R$ is not left-right symmetric.
In the limt of ${M_T \over M_{WR}}=0$, they are given by,
\bea
V^L &=& \left( \begin{array}{ccc}
 1-\frac{\lambda^2}{2} &
\lambda & A \lambda^3 (\rho-i \eta)    \\
-\lambda-i A^2 \lambda^5 \eta  & 1- \frac{\lambda^2}{2} & A \lambda^2   
 \\ 
 A \lambda^3 (1-\rho-i \eta) & -  A 
\lambda^2 - i A^2 \lambda^4 \eta & 1  \\ 
0 & 0 
 & 0
\end{array}
\right),  \\
V^R &=& \left( \begin{array}{ccc}
 1-\frac{\lambda^2}{2} &
\lambda & A \lambda^3 (\rho-i \eta)   \\
-\lambda -i A^2 \lambda^5 \eta & 1- \frac{\lambda^2}{2} & A \lambda^2  
 \\
 0 & 0 & 0 \\
 A \lambda^3 (1-\rho-i \eta) & -  A \lambda^2 -i A \lambda^4 \eta 
 & 1  
\end{array}
\right).
\label{eq:KM0}
\eea
\section{One-Loop Level FCNC}
In the present model, there are many particles
 which may contribute to
the FCNC in the one-loop level.
For example,
there are new contributions to the $\Delta F=2$ transition from box
diagrams involving ordinary
quarks and heavy isosinglet quark intermediate states.
They can contribute to the Feynman diagrams in which  $W_L W_L$,
$W_L W_R$ and $W_R W_R$ are exchanged. $W_R W_R $ exchanged diagrams
are suppressed and we ignore their effect.  The
major contribution from  $W_L W_L$ and $W_R W_L$ exchanged diagrams is
discussed
below.
\\
In the $W_L W_L$ exchanged diagrams, the lightest singlet up type quark 
 $T$ can contribute.  We denote  $M_{12}(SM)$ as the contribution to
the off-diagonal matrix element of neutral meson systems, i.e., $K \bar 
 K $ and $B \bar B$
  in the SM. $M_{12}(LL)$ is the  $W_L W_L$ exchanged
 box
 diagram in the present model in which $T$ quark loop  
 is taken account of.
The deviation from the SM due to $T$ quark loop can be written 
with Inami-Lim functions, 
\bea
{{M_{12}(LL)-M_{12}(SM)} \over M_{12}(SM)}&\simeq&
(\lambda_t^{LL})^2 \biggr( 2 ({S(x_t,x_T) \over S(x_t)}-1 )s_{Lt}^2 c_{Lt}^2 
                  + ({S(x_T) \over S(x_t)}-1 ) s_{Lt}^4 \biggl), \nn \\ 
\lambda_t^{LL}&=&U^{L*}_{t k} U^{L}_{t l},\quad (k,l)=(d,s),(d,b),(s,b),
 \nn \\
x_t&=& ({m_t \over M_{WL}})^2, \quad x_T= ({m_T \over M_{WL}})^2,
\label{eq:LL}
\eea
where $(k,l)=(d,s)$,$(d,b)$,and $(s,b)$ correspond to  $K \bar K$ 
, $B_d \bar B_d$ and $B_s \bar B_s$ mixings respectively. 
$S(x)$ and $S(x,y)$ are the Inami-Lim functions. 
$s_{Lt}$ is the mixing angle of the singlet quark $T$ defined in
Eq.(\ref{eq:sL}). The mixing angle is approximately given by,
\bea
s_{Lt}=O({M_{WL} M_T \over M_{WR}^2}).  
\eea
We find that for $M_T=O(M_{WL})$, the
mixing angle is suppressed as $s_{Lt}=O({M_{WL}^2 \over M_{WR}^2})$.
For $M_T=O(M_{WR})$, we obtain the larger mixing angle,i.e., 
$s_{Lt}=O({M_{WL} \over M_{WR}})$.
When $M_{WR} \simeq 400 \rm (GeV)$, for the former case, the mixing 
angle is $0.04$ and for the latter case, it is $0.2$.
Therefore, for the smaller $M_T=O(M_{WL})$,
 the $T$ quark does not contribute to 
$W_L W_L$ exchanged diagrams at all. 
For the larger mixing angle case, the deviation from the
SM can be as large as $30 \%$  for $M_{WR}\simeq 400 \rm GeV$.
In the subsequent
 analysis, we extensively study the case for the small mixing angle. 
Then  $W_L W_L$ box diagram contribution is not changed from
that of the SM, i.e., $M_{12}(LL) \simeq M_{12}(SM)$. 
For the latter case, we 
need to do more careful analysis and this will be done in the
future publication.\\
When the $T$ quark contribution to $W_L W_L$ exchanged diagrams is small,
 the most important contribution comes from $W_R 
W_L$ exchanged diagrams. \cite{BBS}
In the $W_R W_L$ exchanged diagrams (Fig.1), we must also take account of
both isosinglet and isodoublet quarks
as intermediate states.    
\subsection{$K^0-\bar K^0$ mixing}
The $K^0-\bar K^0$ mixing is one of
the best processes to test the FCNC in our model.
We reexamine the 
 famous enhancement factor in  $W_L-W_R$ box 
diagrams of LR models  \cite{BBS} taking account of both
isosinglet and isodoublet quarks as intermediate states.
The effective Hamiltonian for $W_L-W_R$ exchanged diagram is,
\begin{equation}
  H_{\rm
eff}(LR)=\frac{G_F^2}{4\pi^2}M_{W_L}^2\Lambda_\gamma^{LR}\Lambda_\delta^{RL}
  2 \beta\sqrt{x_\gamma x_\delta} 
  F(x_\gamma,x_\delta,\beta)(\overline d L
s)(\overline d R s) \ ,
\end{equation}
\noindent
where
\begin{equation}
 \Lambda_\gamma^{LR}=V^{L*}_{{q_\gamma}d} V^{R}_{{q_\gamma}s} \ ,
 \qquad \Lambda_\delta^{RL}=V^{R*}_{{q_\delta}d} V^{L}_{{q_\delta}s} \ ,
  \qquad q_{\gamma,\delta}=(u,c,t,U,C,T)  \ ,
\end{equation}
\noindent
and
\begin{equation}
\beta=\left (\frac{M_{W_L}}{M_{W_R}} \right)^2 \ ,\qquad
 F(x_\gamma,x_\delta,\beta)=(4+\beta x_\gamma x_\delta)
I_1 \left(x_\gamma,x_\delta,\beta\right)
-(1+\beta) I_2\left(x_\gamma,x_\delta,\beta\right) \ ,
\end{equation}
\begin{equation}
 x_\gamma \equiv x_{q_i}=\left (\frac{m_{q_i}}{M_{W_L}} \right)^2 \
(\gamma=1\sim 3), \quad
 x_\gamma \equiv X_{Q_i}=\left (\frac{m_{Q_i}}{M_{W_L}} \right)^2 \
(\gamma=4\sim 6) \ ,
\label{parameter}
\end{equation}
\noindent where $i$ 
runs from 1 to 3 and $V^{L/R}$ denotes the $6\times 6$
mixing matrices.
The loop functions $I_1$ and $I_2$ are found to be the same ones given by
Ecker and Grimus\cite{EG}:

\begin{equation}
I_1\left(x_\gamma,x_\delta,\beta\right)
=\frac{ x_{\gamma} \ln x_{\gamma} }
{(1-x_{\gamma})(x_{\gamma}-x_{\delta})(1-\beta x_{\gamma}) }+( x_{\gamma}
\leftrightarrow x_{\delta})-
\frac{\beta \ln \beta }{ (1-\beta x_{\gamma})(1-\beta
x_{\delta})(1-\beta)} ,
\end{equation}
\begin{equation}
I_2\left(x_\gamma,x_\delta,\beta\right)
=\frac{x_{\gamma}^{2}\ln x_{\gamma}}{ (1-x_{\gamma})(1-\beta
x_{\gamma})(x_{\gamma}-x_{\delta}) }
+(x_{\gamma} \leftrightarrow x_{\delta})
-\frac{\ln \beta}{(1-\beta)(1-\beta x_{\gamma})(1-\beta x_{\delta})} .
\end{equation}
\begin{figure}[H]
\begin{center}
\leavevmode\psfig{file=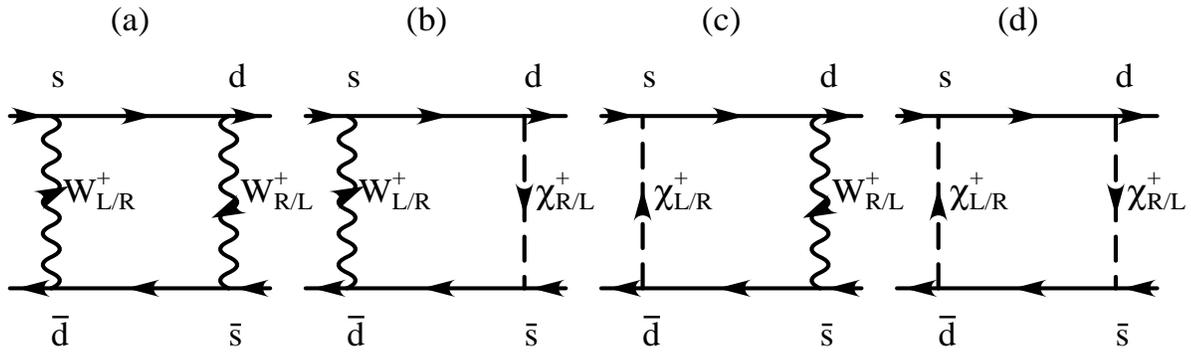,width=17cm}
\caption{Box diagrams for $K^0-\bar K^0$ mixing involving singlet and
doublet quark intermediate states.}
\end{center}
\end{figure}
The contribution to the off diagonal matrix element  $M^K_{12}$ is
\begin{equation}
 M^K_{12}(LR)=
 \frac{G_F^2}{4 \pi^2}M_{W_L}^2\Lambda_\gamma^{LR}\Lambda_\delta^{RL}
  2 \beta\sqrt{x_\gamma x_\delta} F(x_\gamma,x_\delta,\beta)
   \langle  K^0 |(\overline d L s)(\overline d R s)|\overline{K^0}
    \rangle
\frac{1}{2m_K}.
\end{equation}
\noindent
  The $K_L$ and
  $K_S$ mass difference, $\Delta m_K$ and  the $CP$ violating
parameter $\epsilon_K$ are given:

\begin{equation}
 \Delta m_K= 2 \Re M^K_{12} \ , \qquad \ \epsilon_K =\exp (i
 {\pi \over 4}) 
( \frac{\Im M^K_{12}}{\sqrt{2} \Delta m_K} + \xi_0), \quad 
\xi_0={Im A_0 \over Re A_0}. 
\end{equation}
The $LR$ contribution to $\Delta m_K$ is given by
\begin{equation}
 \Delta m_K(LR)=
 \frac{G_F^2}{6 \pi^2}M_{W_L}^2 f_K^2 m_K \kappa
 \sum_{\gamma,\delta=1}^6 \Re(\Lambda_\gamma^{LR}\Lambda_\delta^{RL})
  2 \beta\sqrt{x_\gamma x_\delta} F(x_\gamma,x_\delta,\beta) \ ,
\end{equation}
\noindent
where
\begin{equation}
\langle\overline K^0 |(\overline s L d)(\overline s R d)|K^0\rangle
=\frac{1}{3}\kappa f_K^2 m_K,
\end{equation}
\noindent with
\begin{equation}
\kappa=\frac{3}{4}\frac{m_K^2}{(m_s+m_d)^2}+\frac{1}{8}.
\end{equation}
\noindent
We estimate the matrix element using 
the vacuum-insertion approximation.
In the following estimate of the $LR$ contribution,
 we neglect the effect
of  the
QCD corrections.\par

The sum over $\gamma,\delta=1\sim 6$ can be reorganized in a better way.
By using eq.(\ref{relation}),
 we obtain
 \begin{equation}
 c_{L{q_i}} c_{R{q_i}} \sqrt{x_{q_i}} =- s_{L{q_i}} s_{R{q_i}}
\sqrt{X_{Q_i}}   \ , \quad (i=1,2,3)
\end{equation}
\noindent which lead to a useful formula
\begin{eqnarray}
 \Delta m_K(LR)&=&
 \frac{G_F^2}{6 \pi^2}M_{W_L}^2 f_K^2 m_K \kappa 2 \beta \times \nonumber
\\
 && \sum_{i,j=1}^3 \Re(\lambda_i^{LR}\lambda_j^{RL}) c_{L{q_i}} c_{R{q_i}}
c_{L{q_j}} c_{R{q_j}}\sqrt{x_{q_i} x_{q_j}}
  \tilde F(x_{q_i},x_{q_j},X_{Q_i},X_{Q_j},\beta) \ ,
\label{dm}
\end{eqnarray}
\noindent where
 \begin{equation}
\tilde F(x_{q_i},x_{q_j},X_{Q_i},X_{Q_j},\beta)
=F(x_{q_i},x_{q_j},\beta)-F(X_{Q_i},x_{Q_j},\beta)-F(x_{q_i},X_{Q_j},\beta)
+F(X_{Q_i},X_{Q_j},\beta)\ ,
\label{ftilde}
\end{equation}
\noindent which is a gauge invariant quantity \cite{Gauge}.
$\lambda_i^{LR}$ is defined by
\begin{equation}
 \lambda_i^{LR}=U^{L*}_{{q_id}} U^{R}_{{q_is}}  \ , \qquad
(i=1,2,3).
\end{equation}
Before closing this sub-section, we show the SM contribution
$M^K_{12}(LL)$,
which is used in the following calculations:
 \bea
 M^K_{12}(LL)&=&
 \frac{G_F^2}{12 \pi^2}F_K^2 B_K  m_KM_{W_L}^2 [(\lambda_c^{LL})^2
 \eta_1 S(x_c)+
  (\lambda_t^{LL})^2 \eta_2 S(x_t)
  +2\lambda_c^{LL}\lambda_t^{LL}\eta_3
S(x_c,x_t)], \nn \\
\lambda_i^{LL}&=&U^{L*}_{{q_id}} U^{L}_{{q_is}}  \ , \qquad
(i=1,2,3),
\eea
\noindent where  $S(x)$ and $S(x,y)$ are the Inami-Lim functions, 
 and $\eta_1=1.38$,  $\eta_2=0.57$  and $\eta_3=0.47$ are used.
 The parameter $B_K$ is taken to be $0.75\pm 0.15$ in the following
calculations.

\subsection{Numerical results in LR symmetric limit}
 In the limit of $m_u=0$, the up flavors contribution, i.e., 
 the (i=u and/or j=u) contribution to $\Delta m_K(LR)$
vanishes.
 The loop integrated functions $\tilde F(x_i,x_j,X_i,X_j,\beta)$ for
 charm flavors, top flavors
  and the mixed flavors
    intermediate states are given approximately as:
\bea
\tilde F(x_c,x_c,X_C,X_C,\beta) &\simeq&
 5+ 4 \ln{x_c}- \ln{y_c}+ \ln{\beta}, \nn \\ 
\tilde F(x_t,x_t,X_T,X_T,\beta) &\simeq& 
\frac{4-x_t}{1-x_t}+\frac{x_t^2-2x_t+4}{(1-x_t)^2}\ln{x_t} +\ln{\beta} \ ,
\nn \\
\tilde F(x_c,x_t,X_C,X_T,\beta) &\simeq& 
\frac{4-x_t}{1-x_t}\ln{x_t}+\ln{\beta},
\eea
where $y_i=\beta X_i=m^2_{Q_i}/M^2_{W_R}$.
\noindent
 The contribution of each intermediate state also depends on the
mixings:
\begin{equation}
 \lambda_i^{LR}\lambda_j^{RL}=U^{L*}_{{q_id}} U^{R}_{{q_js}}
U^{R*}_{{q_i}d} U^{L}_{{q_j}s} \ .
\end{equation}
In 
the left-right symmetric limit of the Yukawa couplings, i.e.
 $y_{L(U,D)}=y_{R(U,D)}$ in eq.(\ref{matrix}),
 the relations
$U^{L}_{q_id}=U^{R}_{q_i d}$ and
$U^{L}_{q_is}=U^{R}_{q_i s}$ hold. As we discussed in the
section 5, we can apply the 
Wolfenstein parameterization on $U^L$ and $U^R$.
Taking into consideration these mixing matrices, we can estimate
 contributions of charm flavors, top flavors  
 and mixed flavors of charm and top flavors
 intermediate states. The charm flavor contribution includes
 (c,c), (c,C), (C,c) and (C,C) as intermediate states. The
 top flavor contribution includes (t,t), (t,T), (T,t) and (T,T). 
 The mixed flavor contribution comes from
  (c,t),(C,t),(c,T), and (C,T)
 intermediate states.
 The relative contributions of ordinary quarks and singlet quarks are
found by estimating 
$\tilde F(x_{q_i},x_{q_j},X_{Q_i},X_{Q_j},\beta)$.      
Using the Wolfenstein parameterization,
the mass difference $\Delta m_K(LR)$ is given by, 
 \bea
{\Delta m_K(LR)
\over \Delta m_K}&=& 2 \sqrt2 C_{\epsilon} \kappa \beta 
\{ \lambda^2 x_c \tilde  F(x_c,x_c,X_C,X_C,\beta) \nn \\ &+&
A^4 \lambda^{10} \{(1-\rho)^2-\eta^2 \}
 x_t c_{Rt}^2\tilde F(x_t,x_t,X_T,X_T,\beta) \nn \\
 &+& 
2 A^2 \lambda^6 (1-\rho) \sqrt{x_c x_t} 
 c_{Rt}\tilde F(x_c,x_t,X_C,X_T,\beta) \},\nn \\ 
 C{\epsilon}&=&
{{G_F}^2 {F_K}^2 {m_K} {M_{WL}}^2 \over {6 \sqrt{2} \pi^2 \Delta m_K}}
=3.78 \times 10^4.
 \eea
$x_c \tilde F(x_c,x_c,X_C,X_C,\beta)\simeq -0.015$,
$x_t \tilde F(x_t,x_t,X_T,X_T,\beta)\simeq -14$ and
$\sqrt{x_c x_t}\tilde F(x_c,x_t,X_C,X_T,\beta)\simeq 0.018$ for 
$ M_{WR}=2(\rm{TeV})$.
So, it is found that the charm flavor
contribution is the most important in $\Delta m_K(LR)$
.We also note that 
 the $W_L-W_R$ box graph contribution
  to  $\Delta m_K$ is negative relative to the
$W_L-W_L$ one.
Among the charm flavor contributions, a quarter of the
$\Delta m_K(LR)$ comes from two isosinglet quark intermediate
states, i.e., (C,C). This can be seen that 
  the singlet C quark intermediate states contribution in 
 $\tilde F(x_c,x_c,X_C,X_C,\beta)$ is approximately
 given by  $1 - \ln{y_c}+ \ln{\beta}$.
In the numerical calculation, we must give the mass of the
isosinglet quark C. It is determined as follows.
The mass of charm quark is given by the  seesaw
formulae,
$m_c \cong (\eta_R/m_C) 
y_{RU2} y_{LU2} \eta_L$.
In the left-right symmetric limit, i.e.,
$y_{RU2}\cong y_{RU2}=y_{U2}$, it becomes,
$m_c= 2 ({y_{U2} \over g})^2 {M_{WR} M_{WL} \over m_C}$.
With the assumption of the strength of Yukawa coupling,
$y_{U2}=O(g)$,  we get $m_C=O( 100 \times M_{WR})$. 
In our numerical calculation of $\Delta m_K $ and $\epsilon_K$,
we set
$m_C=100 M_{W_R}$.
The $W_L-W_L$ box
 graph contribution is well known and is given by
 \bea
{\Delta m_K(LL)
\over \Delta m_K}=  {\sqrt 2} C_{\epsilon} B_K
 \lambda^2 x_c \eta_1.
\eea
 The $W_L-W_R$ box graph contribution
  to  $\Delta m_K$ is negative relative to the
$W_L-W_L$ one. Therefore
the experimental value of $K_L$ and $K_S$ mass difference
can be fitted only if there 
is sizable constructive contribution to $\Delta m_K$ due to
the long distance effects or if $W_R$ is sufficiently large 
and $\Delta m_K$ is saturated by the standard model like 
contribution. Including possible long distance effects,
 $\Delta m_K$ consists of three parts.     
\bea
\Delta m_K=\Delta m_K(LL)+ \Delta m_K(LR)+ \Delta m_{Long},
\eea
where $\Delta m_{Long}$ is
 the long distance contribution.
First, we show the  ratio  ${ \Delta m_K(LR) \over \Delta m_K(LL) }$
versus $M_{W_R}$ in Fig.2, where 
 $m_s=120 {\rm MeV}$ or $200 {\rm MeV}$, and  $B_K=0.75$ are used. 
 In order for the $W_L-W_R$ contribution to be  
  smaller than the $W_L-W_L$ 
 one,
$M_{WR}\geq 1.3 {\rm TeV}$ should be satisfied.
In Fig.3, we show the $\Delta m_{Long}$  
 which is needed to explain the
experimental value $\Delta m_K$ in the present model.
If we allow the sizable $\Delta m_{Long}=O(\Delta m_K)$,
 $M_{WR}$ is allowed 
 to be $1 \rm{TeV}$.  
Therefore we conclude the lower bound for
$M_{WR}$ obtained from $\Delta m_K$ is $O(1 \rm{TeV})$,
though
there is large theoretical uncertainty
on the lower bound on $M_{WR}$ from $K_L$
and $K_S$ mass difference.
\begin{figure}[h]
\begin{center}
\leavevmode\psfig{file=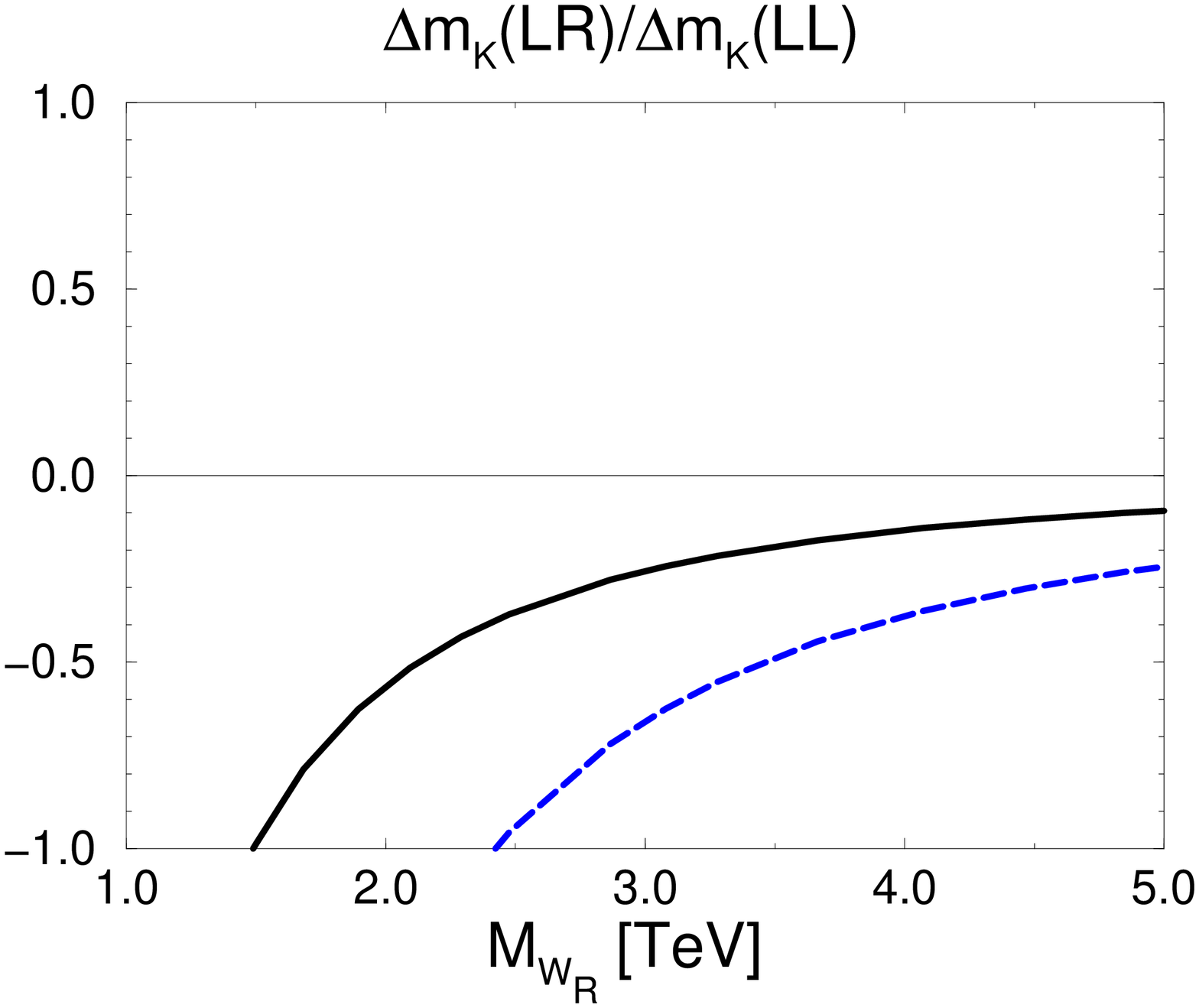,width=16cm,angle=-90}
\caption{ ${\Delta m_K(LR) \over \Delta m_K(LL)} $ as function
of $M_{WR}$. 
The solid line corresponds to $m_s=200(\rm MeV)$ and
the dashed line corresponds to $m_s=120(\rm MeV)$.
We use $B_K=0.75$.}
\end{center}
\end{figure}
\begin{figure}[h]
\begin{center}
\leavevmode\psfig{file=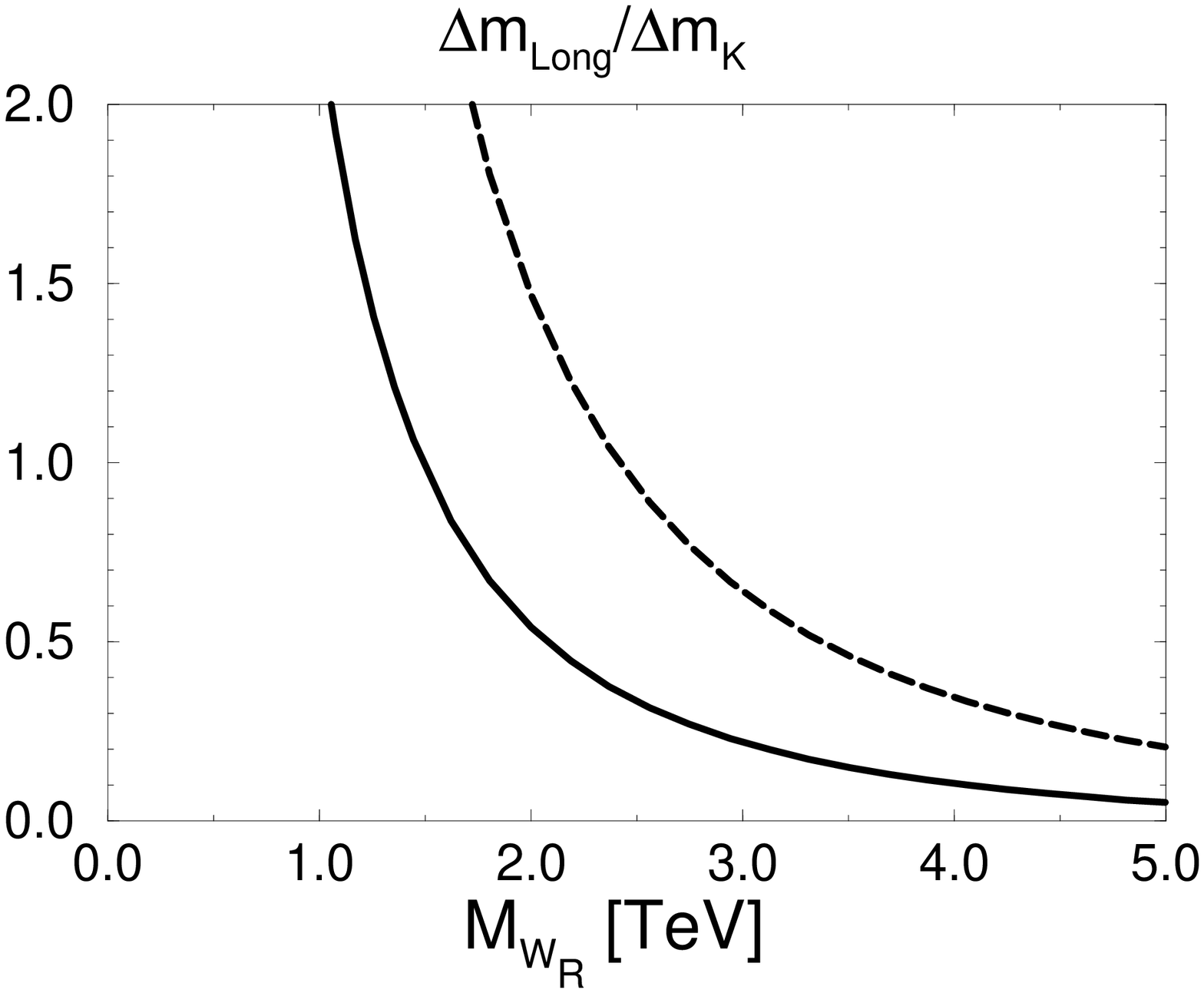,width=16cm,angle=-90}
\caption{ The long distance contribution $\Delta m_{Long}$
which reproduces the $K_L$ and $K_S$ mass difference in the 
present model.
The solid line corresponds to $m_s=200(\rm MeV)$ and
the dashed line corresponds to $m_s=120(\rm MeV)$.
The unit is the experimental value of $\Delta m_K$. We use $B_K=0.75.$}   
\end{center}
\end{figure}
\subsection{$CP$ violation}
Let us discuss the $CP$ violation of the $K^0-\bar K^0$ system. 
By studying  $\epsilon_K$ in the present model,
we can obtain alternative constraint on
 two important parameters $M_{WR}$ and $M_T$ in the model. 
As we discussed in the previous
sub-section, the lower bound on $M_{WR}$ 
is obtained from $\Delta m_K$.
However it is shown that $K_L$ and $K_S$
 mass difference is not sensitive
to  $M_T$ because the top flavors contribution
is tiny in the real part of $M_{12}$.
However, in the imaginary part, the top flavors
contribution becomes important as $M_T$ is larger.
Neglecting  tree level FCNC, there is one  $CP$ violating phase
for the left-handed
mixing matrix $U^L$ and one for the right-handed mixing matrix $U^R$.
In the left-right symmetric limit, 
$U^L=U^R$ holds and  
we have only one $CP$ violating phase.
The $CP$ violating effect of the $W_L-W_R$  exchange 
in the $K^0-\bar K^0$
mixing  is proportional to the $CP$ violating phase.
  The imaginary part of $M^K_{12}$,
which comes from $W_L-W_R$  exchanged diagrams is given as follows:
\begin{eqnarray}
 \Im M^{K}_{12}(LR) &=&
 \frac{G_F^2}{12 \pi^2}M_{W_L}^2 f_K^2 m_K \kappa 2 \beta \times \nonumber
\\
 && \sum_{i,j=1}^3 \Im(\lambda_i^{LR}\lambda_j^{RL}) c_{L{q_i}}
c_{R{q_i}} c_{L{q_j}} c_{R{q_j}}
 \sqrt{x_{q_i} x_{q_j}}\tilde F(x_{q_i},x_{q_j},X_{Q_i},X_{Q_j},\beta) \,
\end{eqnarray}
\noindent
where
\begin{equation}
 \Im(\lambda_c^{LR}\lambda_c^{RL})=
 -2A^2\lambda^6\eta , \quad
 \Im(\lambda_t^{LR}\lambda_t^{RL})=
 2A^4\lambda^{10}\eta (1-\rho) , \quad
 \Im(\lambda_c^{LR}\lambda_t^{RL})=2A^2\lambda^6\eta.
\end{equation}
\noindent
$\epsilon_K$ in the present model can be written as,
\bea
\epsilon_K &=& \epsilon_{LL}+\epsilon_{LR}, \nn \\
\epsilon_{LR}&=& C_{\epsilon} \kappa \beta \eta  
\{-2 A^2 \lambda^6 x_c  \tilde F(x_c,x_c,X_C,X_C,\beta)
+ 2 A^4 \lambda^{10} (1-\rho) 
c_{Rt}^2x_t \tilde F(x_t,x_t,X_T,X_T,\beta) \nn \\
&+& 2 A^2 \lambda^6 \sqrt{x_c x_t} c_{Rt} 
\tilde F(x_c,x_t,X_C,X_T,\beta)\} \exp (i {\pi \over 4}), 
\nn \\
\epsilon_{LL}&=& \frac{1}{2} B_K C_{\epsilon} \eta
\{-2 A^2 \lambda^6 \eta_1 S(x_c,x_c) 
  +2 A^4 \lambda^{10} \eta_2 (1-\rho) S(x_t,x_t) \nn \\ 
  &+& 2 A^2 \lambda^6 \eta_3  S(x_c,x_t) \} \exp(i {\pi \over 4}).
\label{eq:epsilon}
\eea
 For numerical calculation, we must know how $c_{Rt}$ and
$m_T$ depend on the parameters $M_T$ and $M_{WR}$. 
In the present model,
the masses of the top quark and its singlet partner
are given by,
 \begin{eqnarray}
 & m_t& \cong \sqrt{2}\left( \frac{|y_L y_R| }{g^2} \right) M_{W_L}
\frac{1}{ \sqrt {\left(\frac{ |y_R| }{g} \right)^2 +
\left( \frac{M_T}{\sqrt{2}M_{W_R}} \right)^2 } }\ ,\label{mt} \\
 & m_T&  \cong \sqrt{2}M_{W_R} \sqrt{\left(\frac{ |y_R| }{g} \right)^2 +
\left( \frac{M_T}{\sqrt{2}M_{W_R}} \right)^2}\ .
\label{eqn:mass}
\end{eqnarray}
(c.f. Eq.(\ref{eq:8.1})).
In the case of the left-right symmetric limit $y_L=y_R\equiv y$,
the Yukawa coupling $y$ is given in terms of the top quark mass:
\begin{eqnarray}
y^2=g^2\frac{m_t^2}{4M_{W_L}^2}\left (1+\sqrt{1+4\frac{M_{W_L}^2
M_T^2}{M_{W_R}^2 m_t^2}}\right)\ .
\label{Yukawa}
\end{eqnarray}
\noindent
Then the singlet quark mass is given as
\begin{equation}
 m_T=m_t\frac{M_{W_R}}{M_{W_L}}\frac{1}{2}
          \left (1+\sqrt{1+4\frac{M_{W_L}^2 M_T^2}{M_{W_R}^2
m_t^2}}\right)\ .
\end{equation}
\noindent 
In the case of $M_{W_R}\gg M_T$, we get
$m_T \simeq m_t{M_{W_R}}/{M_{W_L}}$.
Thus, the physical mass of the singlet
 $T$ quark is determined if
$M_{W_R}$ and $M_T$ are fixed.
In Fig.4, we show the magnitude of  Yukawa coupling
 in the region of $M_T\leq M_{W_R}$.  The Yukawa coupling
 increases as $M_T$ is larger.
 Increasing  $M_T$ further may invalidate 
 the perturbative calculation.
   In Fig.5, $c_{Rt}$ is shown versus $M_T/M_{W_R}$.
As far as $M_T\ll M_{W_R}$, 
the singlet-doublet mixing $c_{Rt}$ is approximately given
by $c_{Rt}\sim M_T/M_{W_R}$.
Coming back to  $\epsilon_K$, the top flavor contribution
is suppressed by the factor. Such suppression mechanism
is distinctive in the present left-right model. 
$c_{Rt}$ increases as $M_T$ is larger.
So the top flavor contribution is 
more important as $M_T$ increases.\\ 
\begin{figure}[H]
\begin{center}
\leavevmode\psfig{file=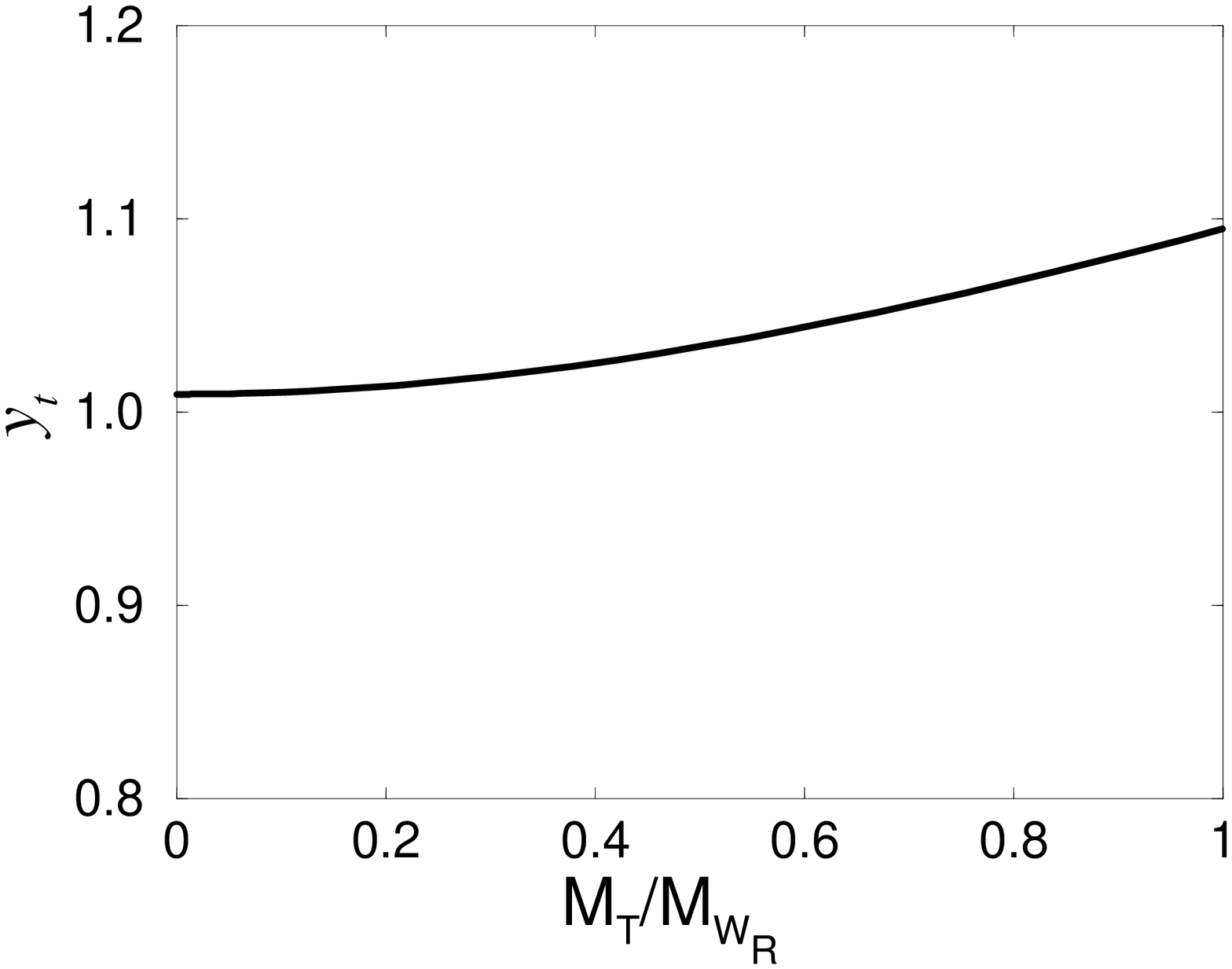,width=14cm,angle=-90}
\caption{ The Yukawa coupling $y_t$ 
as function of ${M_T \over M_{WR}}$.}
\end{center}
\end{figure}
\begin{figure}[H]
\begin{center}
\leavevmode\psfig{file=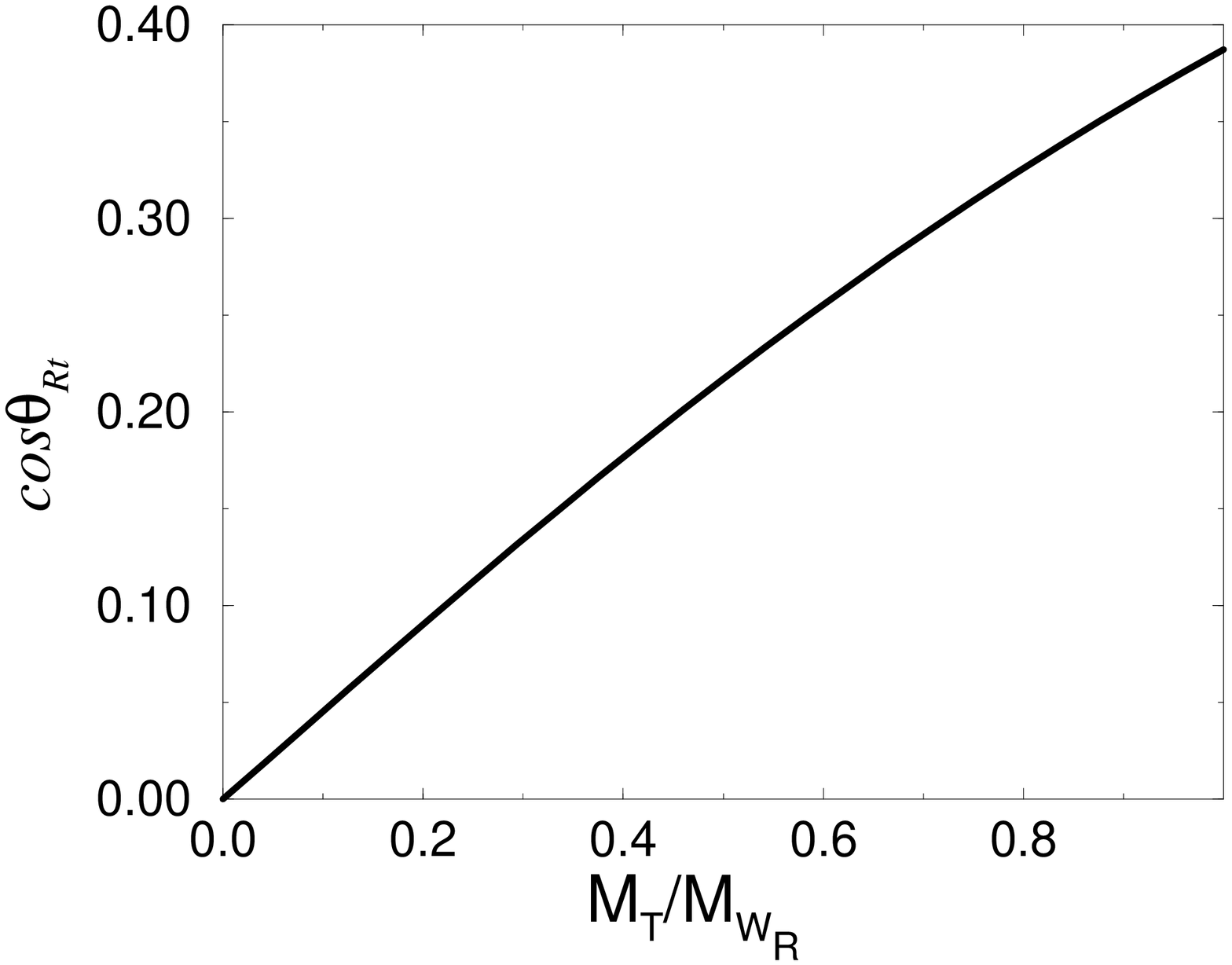,width=18cm,angle=-90}
\caption{The singlet and doublet mixing angle 
$\cos\theta_{Rt}$ as function of ${M_T \over M_{WR}}$.}
\end{center}
\end{figure}
There are two distinctive cases on the contribution to $\epsilon$
 parameters. One  corresponds to the case that ${ \epsilon_{LR} 
 \over \epsilon_{LL}} $ is positive and the other
 corresponds to $ {\epsilon_{LR} \over \epsilon_{LL}}$
 is negative. 
 The ratio is shown as a function of $M_T$ in Fig.6,
  where $M_{W_R}=0.5$, $1.5$, and
  $3 \ {\rm TeV}$ are taken. The ratio does not depend on $\eta$
  and we set $\rho=0$.
  As seen in this figure, 
  the $LR$ contribution is negative in $M_T\geq 60
{\rm GeV}$,
   but positive in $M_T\leq 60 {\rm GeV}$ relative to $LL$
   contribution. (See Eq.(\ref{eq:epsilon}).)
\begin{figure}[H]
\begin{center}
\leavevmode\psfig{file=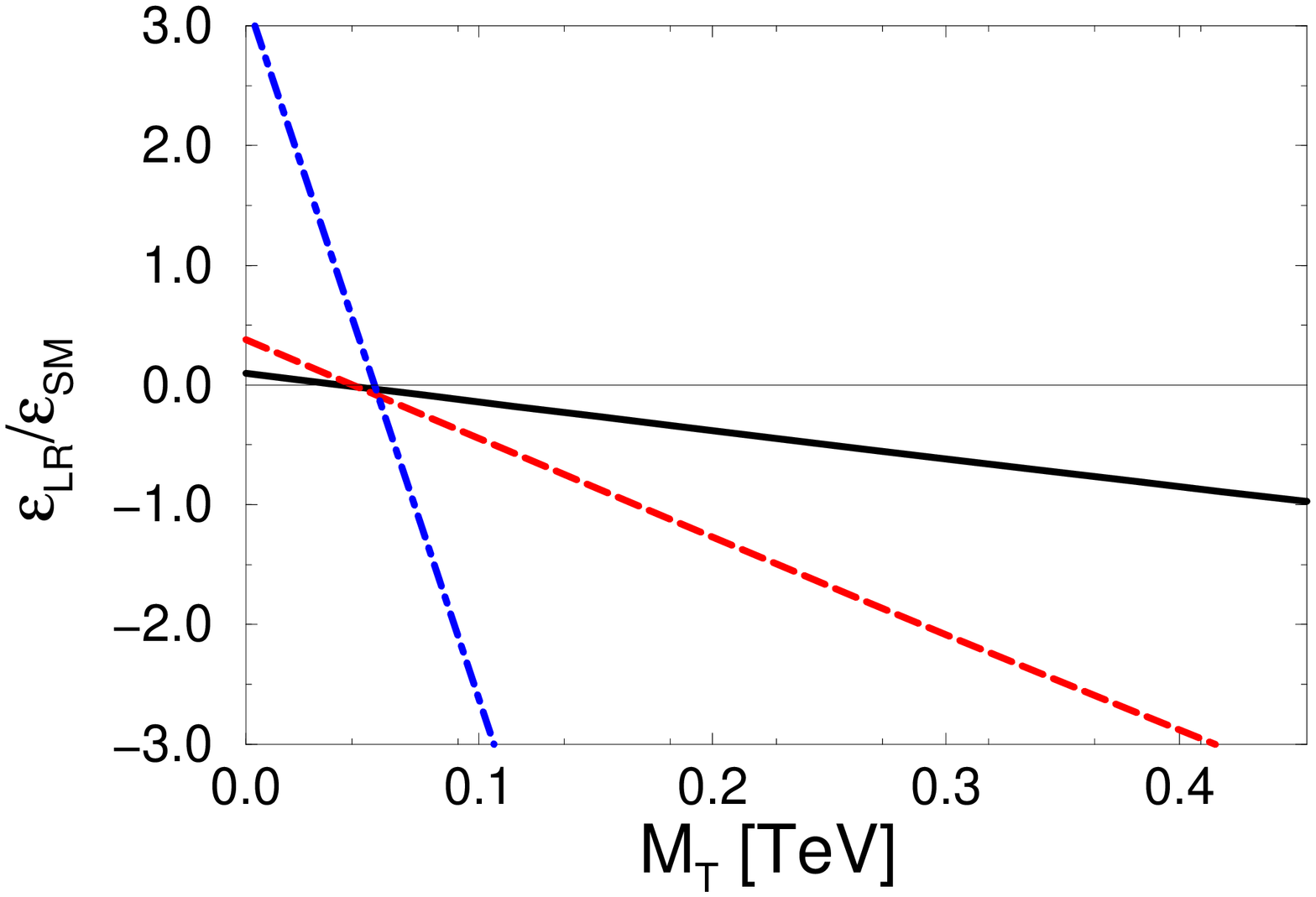,width=18cm,angle=-90}
\caption{${\epsilon_{LR} \over \epsilon_{LL}}$
as function of $M_T$.
The solid line corresponds to $M_{WR}=3(\rm TeV)$,
the dashed line corresponds to $M_{WR}=1.5(\rm TeV)$
and the dotted-dashed line corresponds to $M_{WR}=0.5(\rm TeV)$.   
$\rho$ is set to be zero. $m_s=160(\rm MeV)$ is used.
}
\end{center}
\end{figure}
If $\epsilon_{LR} \ne 0$ and
we would use the standard
model expression to fit $\epsilon_K$,
i.e., with  
 $\epsilon_K=\epsilon_{LL} (\rho,\eta)$,
there would  be disagreement  between 
the allowed region of $(\rho,\eta)$ determined by B
physics data only and that  
obtained by $\epsilon_K$. 
Specifically, here we mean $|V_{ub}|$, $|V_{cb}|$ from semileptonic 
B decays and $|V_{td}|$ from
$B \bar B$ mixing as B physics data.  When we obtain the
constraint on $|V_{td}|$ from the $B \bar B$ mixing,
we can use the SM expression for $M_{12}$ because 
the deviation from the SM in $W_L W_L$ exchanged diagrams
are negligible for the small mixing angle $s_{Lt}=O \bigl(({M_{WL} \over
M_{WR}})^2 \bigr)$
 and $W_R W_L$ exchanged diagrams do not contribute to 
$B \bar B$ mixing as we show later. After all, for the small
mixing angle case, 
 the B physics data is not affected by the presence of the
 new physics.
Then the mismatch between the constraint of $\epsilon_K$
and that of the B physics data  will be manifested in the following ways.
\begin{itemize}
\item[(1)] {The  ($\rho ,\eta$) determined by the B
physics data is above the line obtained from the
constraint  $\epsilon_K=\epsilon_{LL}(\rho,\eta)$.
This case corresponds to $\epsilon_{LR}/\epsilon_{LL} <0 $.
}
\item[(2)] {The  ($\rho ,\eta$) determined by the B
physics data is below the line obtained from the
constraint  $\epsilon_K=\epsilon_{LL}(\rho,\eta)$.
This case corresponds to $ \epsilon_{LR}/\epsilon_{LL} >0 $.
}
\end{itemize}
Qualitatively, this mismatch is
understood as follows.  
 If the $CP$
violating parameter $\eta$ determined by the  B
physics data is larger than the one determined by 
the standard model constraint $\epsilon_K=\epsilon_{LL}$,
we need some negative contribution to $\epsilon_K$ relative to
$\epsilon_{LL}$. If the $\eta$ determined by the  B physics data
is smaller, then the $\epsilon_{LL}$ is not enough to explain
$\epsilon_K$ and we need positive contribution to $\epsilon_K$.
The present data of $B_d \bar B_d$ mixing and $|V_{ub}|$
as well as the bag parameters $B_K$ and $B_B$
are not sufficiently accurate to discriminate the one case from 
the others. 
Of course, this mismatch may disappear by using the
full expression for $\epsilon_K$ in the present model
with the parameters  $M_T$ and $M_{WR}$ suitably adjusted.
In Fig.7, we show the allowed region for 
($M_{WR}, {M_T \over M_{WR}}$)
using the full expression for $\epsilon_K$,
\bea
\epsilon_K = \epsilon_{LL} (\rho, \eta)
 + \epsilon_{LR} (\rho,\eta, M_{WR}
 ,M_T). 
\eea  
Taking into account of
 the present allowed region for ($\rho, \eta$)
 obtained from $B_d \bar{B_d}$ mixing, $|V_{cb}|$ 
 $|V_{ub}|$ and $\epsilon_K$,
 the allowed region for 
 ($M_{WR}, {M_T \over M_{WR}}$) is plotted.
\begin{figure}[H]
\begin{center}
\leavevmode\psfig{file=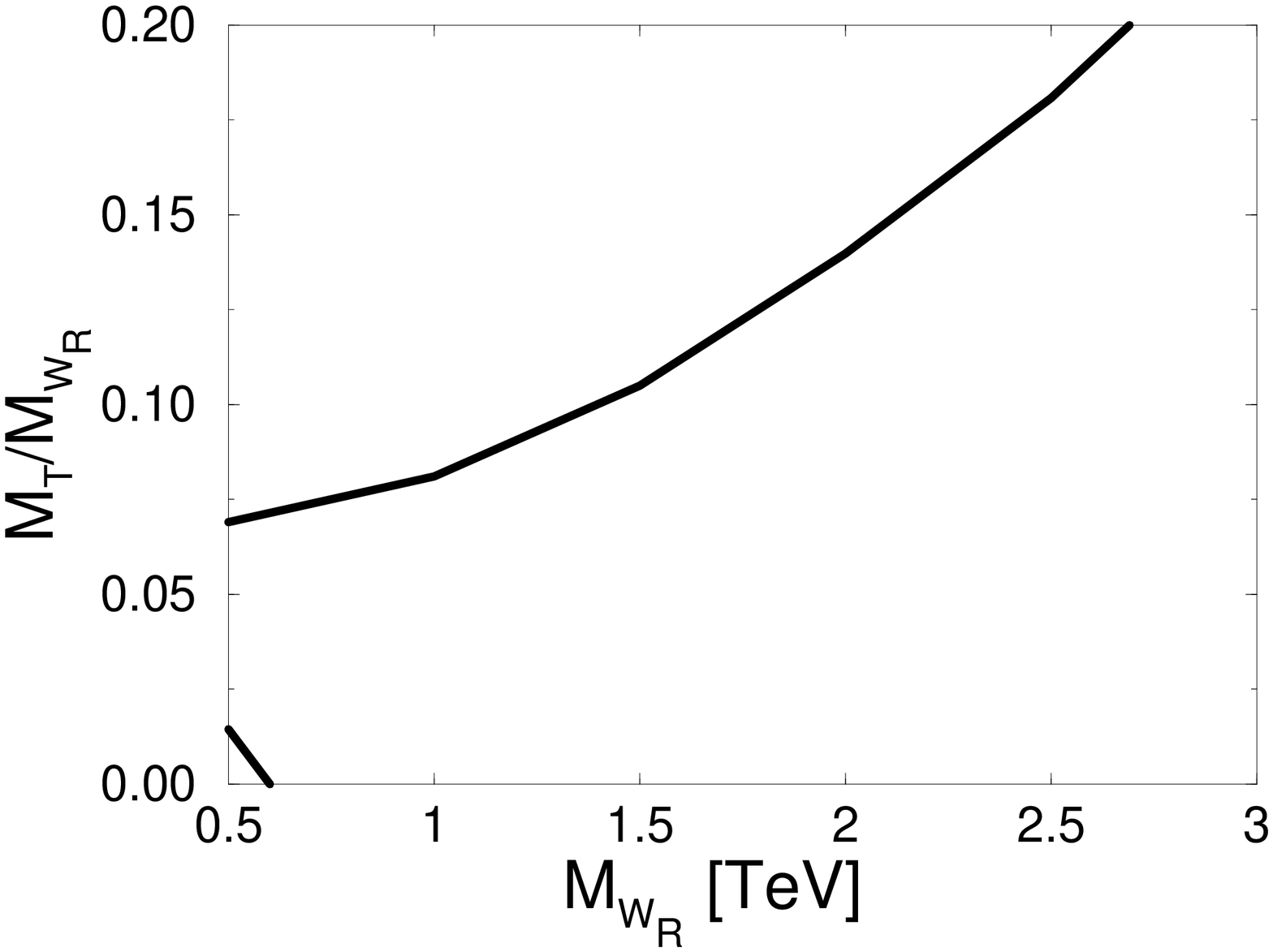,width=18cm,angle=-90}
\caption{The allowed region for  $(M_{WR},
{M_T \over M_{WR}})$. The region between the upper curve
and the lower curve is allowed. We take account of the
present constraint for
$(\rho, \eta)$  obtained from $B_d \bar{B_d}$ mixing
and $|V_{ub}|$. $B_K=0.75$ and $m_s=160 ({\rm MeV})$ are
 used. The upper curve corresponds to $(\rho,\eta)=(-0.087,0.43)$
and the lower curve corresponds to $(\rho,\eta)=(0.25,0.14)$.  
}
\end{center}
\end{figure}
 Depending on the two cases, the allowed
regions of ($M_{WR}, {M_T \over M_{WR}}$)
are quite different from each others.
  The lower curve corresponds to the case  (2), 
  i.e., $ {\epsilon_{LR}
  \over \epsilon_{LL}} >0$. 
 The upper bound on 
$M_{WR}$ is
obtained in this case.
 The upper curve corresponds to the case (1), i.e.,
  $  {\epsilon_{LR} \over \epsilon_{LL}} <0$.
  Since $\eta$ is larger than one which would be
  obtained by the SM fit, the negative contribution is needed.
  This is achieved by taking 
  ${M_T \over MR}$ larger. To maintain this effect while taking $M_{WR}$
  larger, ${M_T \over M_{WR}}$ must increase. 
   We use the following  experimental inputs to determine the
   allowed range of $(\rho, \eta)$:
    $|V^{\rm CKM}_{cb}|=0.0395\pm 0.0017$, 
    $m_t=175.6\pm 5.5 {\rm GeV}$,
	$|V^{\rm CKM}_{ub}/V^{\rm CKM}_{cb}|=0.08\pm 0.016$ and
    $|V^{\rm CKM}_{td}|=0.0084\pm 0.0018$ 
	\cite{PDG}\cite{Buras}.
  We also use $B_K=0.75$.    The above constraints
are shown in 
 $(\rho, \eta)$ plane. (See Fig.9.)
 In order to find the $B_K$ dependence of our result, we also show
  the allowed region  by taking $B_K=0.6\sim 0.9$ with  
  $\rho=0$ and $\eta=0.35$
   in Fig.8.
\begin{figure}[H]
\begin{center}
\leavevmode\psfig{file=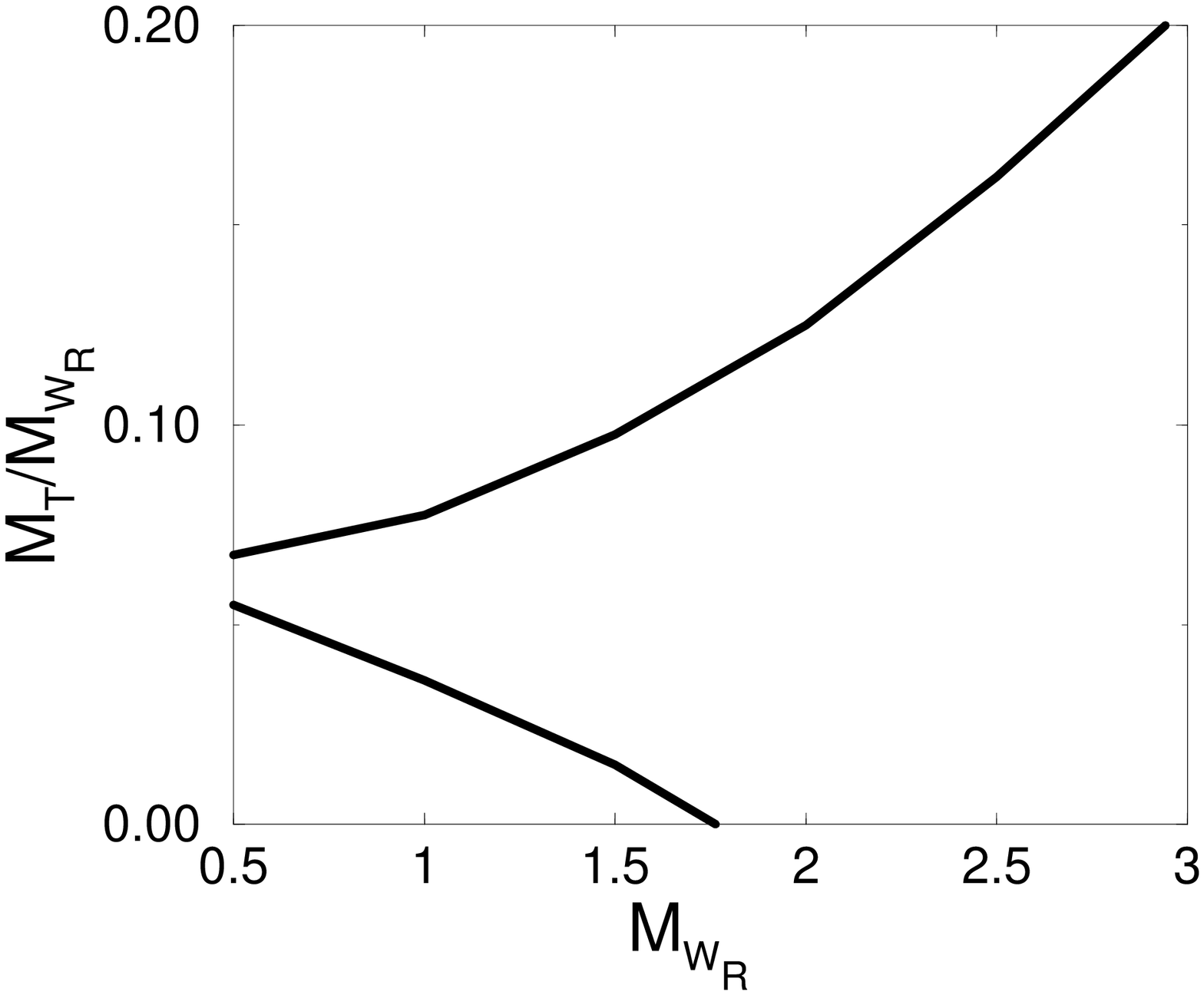,width=18cm,angle=-90}
\caption{ The allowed region for 
$(M_{WR},
{M_T \over M_{WR}})$ with $(\rho, \eta)=(0, 0.35)$.
The lower curve corresponds to $B_K =0.6$ and the
upper curve corresponds to $B_K=0.9$. The region between
two curves is allowed. $m_s=160 ({\rm MeV})$ is used.
}
\end{center}
\end{figure}
%
\begin{figure}[H]
\begin{center}
\leavevmode\psfig{file=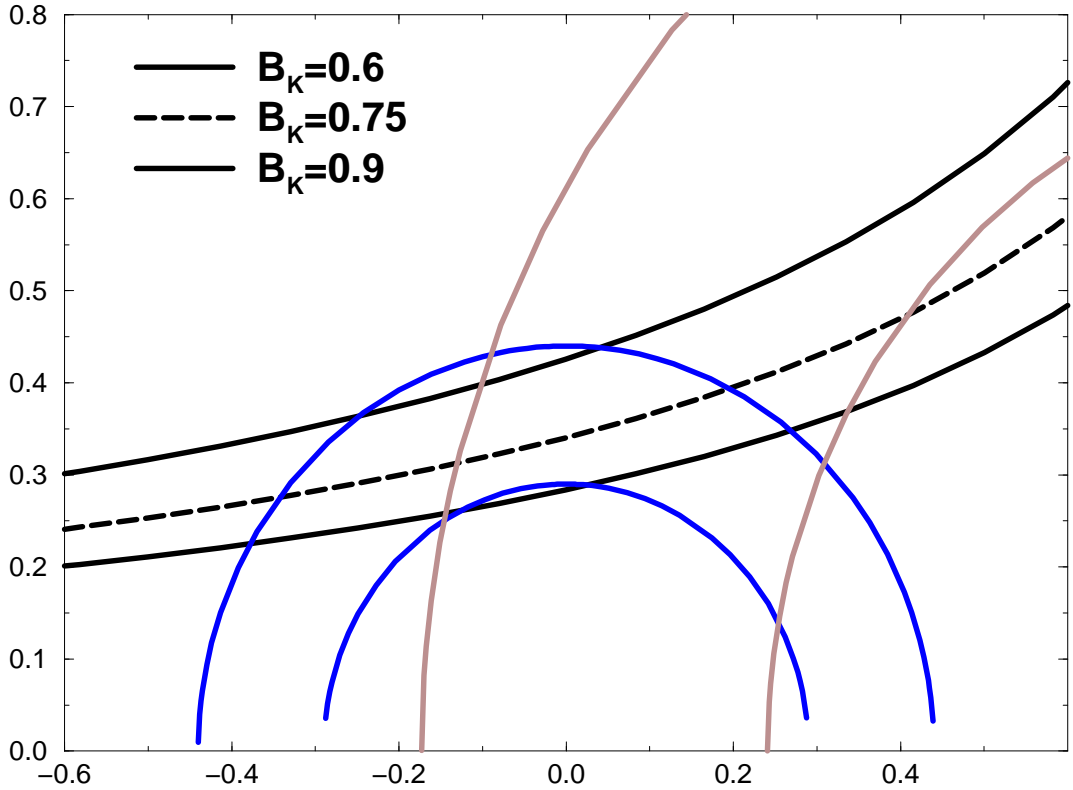,width=18cm,angle=-90}
\caption{ The allowed region in
$(\rho, \eta)$ plane obtained from $|V_{ub}|$,
 $|V_{cb}|$
and $B_d \bar B_d$ mixing. To extract $|V_{td}|$ from
$B_d \bar B_d$ mixing, we use the SM expression for $x_d$.
This is a good approximation for the small angle case, i.e., 
$s_{Lt}=O\biggl( ({M_{WL} \over M_{WR}})^2 \biggr)$.
 (See also the text below
Eq.(\ref{eq:LL}).)         
}
\end{center}
\end{figure}
We can also discuss  the $W_L-W_R$  exchange effect on $B^0_d-\overline
B^0_d$ mixing.
The  contribution to $\Delta m_{B^0_d}$ is given as
\begin{eqnarray}
 \Delta m_{B^0_d}(LR) &=&
 \frac{G_F^2}{6 \pi^2}M_{W_L}^2 f_{B_d}^2 m_B \kappa 2 \beta \times
\nonumber \\
 && \sum_{i,j=1}^3 |\lambda_i^{LR}\lambda_j^{RL}| c_{L{q_i}} c_{R{q_i}}
c_{L{q_j}} c_{R{q_j}}
  \sqrt{x_{q_i} x_{q_j}}
  \tilde F(x_{q_i},x_{q_j},X_{Q_i},X_{Q_j},\beta) \ ,
\end{eqnarray}
\noindent
where $\kappa\simeq 3/4$ and
\begin{equation}
 \lambda_i^{LR}\lambda_j^{RL}=U^{L*}_{{q_id}} U^{R}_{{q_jb}}
U^{R*}_{{q_i}d} U^{L}_{{q_j}b} \ .
\end{equation}
\noindent
In the SM, the top quark intermediate state dominates the
$B^0_d-\overline B^0_d$ mixing.
 The $W_L-W_R$  exchange contribution  is also dominated by the top
flavor intermediate state,
however, this contribution is suppressed by $c_{Rt}$,
that is of order $10^{-4}$ compared with the SM contribution.
 Thus, for $B^0_d-\overline B^0_d$ mixing,
 the $LR$ effect is negligible.
\section{Rare Decays of $K$ Mesons}

Experiments in the $K$ meson system have entered a new period with the
observation of the rare process $K^+ \rightarrow
\pi^{+}\nu \bar{\nu}$, and the dedicated search for $K_{L} \rightarrow
\pi^{0} \nu \bar{\nu}$. 
Recently, the signature of the decay $K^+ \rightarrow \pi^+  \nu
\bar{\nu}$ has been observed by E787 Collaboration \cite{E787} 
and the reported branching ratio is $4.2^{+9.7}_{-3.5}\times 10^{-10}$, 
which is consistent with the value predicted by the SM. 
Additional (and improved) data are expected 
in the near future. In view of this situation, a detailed study of the 
rare $K$ meson decays is necessary. 
The decay $K_L \rightarrow \pi^0 \nu \bar{\nu}$ is one of the most
promising processes, since it is  a $CP$ violating mode in the SM. 
This mode is theoretically clean to extract the CKM 
parameter $\eta$ \cite{BURA}.  
We  investigate rare $K$ meson decays 
in the present  model  introducing right handed 
neutrinos. However, in the model, the neutrino masses are zero in
the tree level and lepton flavor is well conserved 
( see analyses within other models \cite{SEE}).
Scalar and tensor operators 
appear due to the LR box diagrams, 
in which both the left and right handed gauge bosons, $W_{L}$ and 
$W_{R}$, are exchanged. 
The scalar operators have an enhancement factor $M_{K}/m_{s}$ 
in the matrix element $<\pi|\bar{s}d|K>$.
Thus, the scalar operator may make a large contribution to the rare 
$K$ meson decays, $K^+ \rightarrow \pi^+ \nu \bar{\nu}$ 
and $K_{L,S} \rightarrow \pi^0 \nu \bar{\nu}$. 
An important point is that the $CP$ property of the scalar interaction 
is different from the V-A interaction 
$(\bar{s}d)_{V-A}(\bar{\nu}_{l}\nu_{l})_{V-A}$ in the SM.
The decay $K_L \rightarrow \pi^0 \nu \bar{\nu}$ through 
the scalar operator is not a $CP$ violating one, so we have a non-zero 
branching ratio $B(K_L \rightarrow \pi^0 \nu_{l} \bar{\nu}_{l})$ even in 
the $CP$ conserved limit ($\eta \rightarrow 0$).
Thus, it is important to estimate the size of the effect 
of the scalar operator on the  pion energy spectrum. 
\subsection{Rare decays  by   scalar operator}

The rare $K \rightarrow \pi \nu \bar{\nu}$ decays are 
loop-induced FCNC processes in the SM, that, being  
dominated by short-distance physics, are theoretically very clean modes 
\cite{BURA}.
The matrix elements involved in these decays are related to the
experimentally well known decay $K^+ \rightarrow \pi^0 e^+ \nu$
using isospin symmetry, and  corrections to this relation have been 
studied \cite{MARC}. We start our analysis of such decays with the
following effective Lagrangian, which is produced from the LR box
diagram in addition to the SM contributions:
\begin{eqnarray}
{\cal L}_{eff}&=&
-
\frac{4 \kappa G_{F}}{\sqrt{2}}
\sum_{l=e,\mu ,\tau}
\left[
~~C_{SM}^{l}(\bar{s}_{L}\gamma^\mu d_{L})
(\bar{\nu}_{L,l}\gamma_{\mu}\nu_{L,l})
\right.
\nonumber \\
& & \hspace{1cm}
\left.
+S_{LR}^{l}(\bar{s}_{L} d_{R})(\bar{\nu}_{L,l}\nu_{R,l})
+S_{RL}^{l}(\bar{s}_{R} d_{L})(\bar{\nu}_{R,l}\nu_{L,l})
\right.
\nonumber \\
& & \hspace{1cm}
\left.
+T_{LR}^{l}(\bar{s}_{L}\sigma^{\mu\nu}
d_{R})(\bar{\nu}_{L,l}\sigma_{\mu\nu}\nu_{R,l})
+T_{RL}^{l}(\bar{s}_{R}\sigma^{\mu\nu}
d_{L})(\bar{\nu}_{R,l}\sigma_{\mu\nu}\nu_{L,l})
+ h.c.
\right],
\end{eqnarray}
where $\kappa= \alpha/(2 \pi \sin^2\Theta_{W})$.
The first term is the SM contribution \cite{BURA,INAM}, and from the second
to the fifth are the new contributions arising in this model. 
The scalar and tensor operators generally appear from box
diagrams when one considers a model which contains the right-handed
charged gauge boson $W_{R}$. As a concrete analysis, in section 7.3,
we investigate the pion energy spectrum for LR symmetric
scenario,
$S^l_{LR} = S^l_{RL}$ and $T^l_{LR} = T^l_{RL}$, which correspond to the
limit 
$ \cos\theta_{Ls}=\cos\theta_{Rd}=1$ with $U^{R}= U^{L}$.

There are also penguin diagram
contributions to the process we are interested in.  
However, only the box diagrams produce the scalar operator
$(\bar{s}d)_{S}(\bar{\nu}\nu)_{S}$ in the effective Lagrangian, 
thus, we do not consider the contributions from penguin diagrams 
in this paper. 

First we show the decay amplitudes for the neutral K meson states 
$K_{L}$ and $K_{S}$
\footnote{We use a conventional phase choice where
$ |K_{L,S}>\equiv p |K^0> \pm q 
|\bar{K}^{0}>, ~CP |K^0> \equiv - |\bar{K}^{0}>$.}.
The decay amplitudes $A(K_{L,S}\rightarrow\pi^0 \bar{\nu}\nu)$ are:
\begin{eqnarray}
A(K_{L,S}\rightarrow \pi^0 \bar{\nu}_{l}\nu_{l})
&=&
-\frac{G_{F}\kappa}{\sqrt{2}}
\left(
\left( p C_{SM}^{l} \mp q C_{SM}^{l \ast} \right)
<\bar{s}\gamma^{\mu}d>
\left( \bar{\nu}_{l}\gamma_{\mu}(1-\gamma_{5})\nu_{l} \right)
\right.
\nonumber \\
& &
\left.
+
\left(p( S_{LR}^l+S_{RL}^l)\pm q(S_{LR}^l+S_{RL}^l)^{\ast}\right)
<\bar{s}d> (\bar{\nu}_{l}\nu_l)
\right.
\nonumber \\
& & 
\left.
+
\left(
p (S_{LR}^l -S_{RL}^l)\mp q (S_{LR}^l - S_{RL}^l )^{\ast}
\right)
<\bar{s}d> (\bar{\nu}_{l}\gamma_{5}\nu_{l})
\right.
\nonumber \\
& &
\left.
+
4 \left(
p (T_{LR}^l+T_{RL}^l)\mp q (T_{LR}^l+T_{RL}^l)^{\ast}
\right) <\bar{s}\sigma^{\mu\nu}d> (\bar{\nu}_{l}\sigma_{\mu\nu}\nu_{l})
\right.
\nonumber \\
& &
\left.
+
4 \left(
p (T_{LR}^l-T_{RL}^l)\pm q (T_{LR}^l-T_{RL}^l)^{\ast}
\right) <\bar{s}\sigma^{\mu\nu}d>
(\bar{\nu}_{l}\sigma_{\mu\nu}\gamma_{5}\nu_{l})
\right),
\end{eqnarray}

The $CP$ conserved limit corresponds to $p=q$, with all 
coefficients $C_{SM}^l, S^l_{LR}, S^l_{RL}, T^l_{LR}$ and $T^l_{RL}$ real. 
In this limit, the decay amplitude $A(K_L \rightarrow \pi^0
\bar{\nu}_{L}\nu_{L})$ 
through the V-A interaction is zero, while 
$A(K_S \rightarrow \pi^0 \bar{\nu}_{L}\nu_{L})$ is nonzero, 
and the decays through the scalar operators $A(K_{L,S}\rightarrow 
\pi^0 \bar{\nu}_{R}\nu_{L},\pi^0\bar{\nu}_{L}\nu_{R})$ 
remain non-zero generally.
In the LR symmetric parameterization, $K_S$ decay through 
the scalar operators is the $CP$ violating mode, while $CP$ is conserved
for
$K_L$ decay. 
Thus decays of neutral $K$ meson in the LR symmetric parameterization
are summarized as follows:
\begin{eqnarray}
&&
\bullet
K_{L} \mbox{decay}~ 
\left\{
\begin{array}{lll}
(\bar{s}d)_{V-A} (\bar{\nu}_l\nu_l)_{V-A} &\Rightarrow& CP \hspace{-15pt}~/
,\\
(\bar{s}d)_{S}(\bar{\nu}_l \nu_l)_{S}&\Rightarrow&CP ~\mbox{Conserving\ }
,\\
(\bar{s}d)_{T}(\bar{\nu}_l \nu_l)_{T} &\Rightarrow& CP \hspace{-15pt}~/, 
\end{array}
\right .
\nonumber \\
%
%
&&
\bullet
K_{S} ~\mbox{decay}~ 
\left\{
\begin{array}{lll}
(\bar{s}d)_{V-A} (\bar{\nu}_l\nu_l)_{V-A}&\Rightarrow&CP ~\mbox{Conserving\
,} \\
(\bar{s}d)_{S} (\bar{\nu}_l\nu_l)_{S}&\Rightarrow&CP \hspace{-15pt}~/, \\  
(\bar{s}d)_{T}(\bar{\nu}_l \nu_l)_{T}&\Rightarrow&CP ~\mbox{Conserving\ .}
\end{array}
\right .
\nonumber
\end{eqnarray}

Experimentally we do not observe the neutrinos, and the pion 
energy spectrum is obtained by summing these contributions 
which have different $CP$ properties with each other. 
The $K_{L}$ decay through the V-A operator is suppressed due to $CP$
symmetry, 
but decays 
through the scalar operators are $CP$ conserving ones, furthermore 
their matrix elements are enhanced,
$<\pi^0|\bar{s}d|K^0>\sim\frac{M_K^2}{m_{s}}f^{\pm}$ 
( where $<(\bar{s}d)_{V-A}>\equiv f_{+} p_{+}^{\mu}+f_{-} p_{-}^{\mu}, 
p_{\pm}=P_{K}\pm p_{\pi}$ ), 
as we can see from the equation of motion in the next subsection.
Thus, the contribution of the scalar interaction to 
the decay amplitude $A(K_{L} \rightarrow \pi^0 \nu \bar{\nu})$
is sizable and dominates in the $CP$ conserving limit.
The decay amplitude for the charged $K$ meson, $A(K^{+} \rightarrow \pi^{+}
\nu \bar{\nu})$, is obtained in the same way:
\begin{eqnarray}
A(K^{+} \rightarrow \pi^+ \bar{\nu}_{l}\nu_{l})
&=&
-\frac{G_{F}}{\sqrt{2}}\kappa 
\left(~
 C_{SM}
<\bar{s}\gamma^{\mu}d> (\bar{\nu}_{l}\gamma_{\mu}(1-\gamma_{5})\nu_{l})
\right.
\nonumber \\
& & 
\left.
+ \left(S_{LR}+S_{RL}\right)
<\bar{s}d> (\bar{\nu}_{l}\nu_{l})
\right.
\nonumber \\
& & 
\left.
+\left(S_{LR}-S_{RL}\right)
<\bar{s}d> (\bar{\nu}_{l}\gamma_{5}\nu_{l})~
\right.
\nonumber \\
& & 
\left.
+2\left(T_{LR}+T_{RL}\right)
<\bar{s}\sigma^{\mu\nu}d> 
(\bar{\nu}_{l}\sigma_{\mu\nu}\nu_{l})
\right.
\nonumber \\
& & 
\left.
+2\left(T_{LR}-T_{RL}\right)
<\bar{s}\sigma^{\mu\nu}d> 
(\bar{\nu}_{l}\sigma_{\mu\nu}\gamma_{5}\nu_{l})
\right)
\label{eqn:amp},
\end{eqnarray}
where $<{\cal O}>=<\pi^+|{\cal O}|K^+>$.

\subsection{The matrix elements and the coefficient 
functions in the LR model}
%
In this section, we explain our estimations of the matrix elements 
and show explicit forms of the coefficient functions in the  LR model.
The matrix elements in the SM, $<(\bar{s}d)_{V-A}>$, can be related to
the matrix elements of the experimentally well known leading decay 
$K^{+} \rightarrow \pi^{0} e^{+} \nu $ using isospin symmetry. 
So the form factors 
$f_{\pm}^{K^+\rightarrow\pi^+ \nu \bar{\nu}}$ and
$f_{\pm}^{K^0\rightarrow\pi^0 \nu \bar{\nu}}$ are written in terms of 
$f_{\pm}^{K^0\rightarrow\pi^0 e^+ \nu}$.
The matrix element of the scalar operator $<(\bar{s}d)_{S}>$
is also related to these by the equation of motion:
\begin{eqnarray}
<\pi|\bar{s}d|K>
&=&
\frac{p_{-}\cdot<\pi|(\bar{s}d)_{V-A}|K>
      }{m_{d}-m_{s}}
\nonumber \\
&=& 
f_{+}\frac{M_K^2-M_{\pi}^2}{m_d-m_s}
+
f_{-}\frac{p_{-}^2}{m_d-m_s}, 
\end{eqnarray}
where $f_{\pm}$ are the form factors of the corresponding matrix element 
$<\pi|(\bar{s}d)_{V-A}|K>$. 
We estimate the matrix element of the tensor operator
using the NJL model. First we note that the matrix element of
the operator is the second order of the typical momentum.
Therefore in the sense of the chiral perturbation , it is enough to
estimate the form factor $f_T$ at zero momentum transfer,
which is defined by 
$<\bar{s}\sigma^{\mu \nu}d >\equiv \frac{f_T}{M_{K}} \cdot
\left(p_{+}^{\mu}p_{-}^{\nu}-p_{-}^{\mu}p_{+}^{\nu}\right)$.
the SU(3) breaking effect may be also neglected.
Calculating the Feynman diagram shown in Fig.10, the form
factor is given by
\begin{eqnarray}
f_T
&=&
\frac{-i N_{c}}{4 \sqrt{2} \pi^2}
\frac{m_{q} M_K}{f_{\pi}^2},
\end{eqnarray}
where $m_q$ is a constituent quark mass and $f_\pi$ is the
pion decay constant. As a rough order estimate,
we assume $m_q=200 $ MeV, $f_\pi =120$ MeV.
Then we obtain , $f_T=-0.37 i$.
Substituting the value in the pion energy spectrum, we
can see that it is safely neglected.
\begin{figure}[H]
\begin{center}
\leavevmode\psfig{file=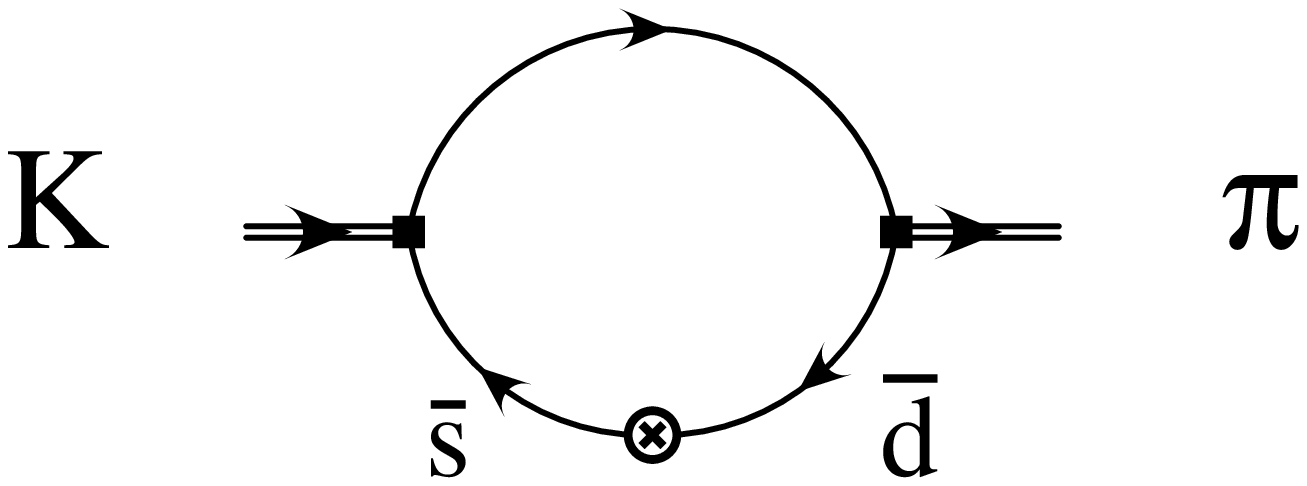,width=8cm}
\caption{The Feynman diagram for the matrix element of the
{\it tensor} operator in NJL model.
$\otimes$ denotes an insertion of the tensor operator 
${\bar s} \sigma_{\mu \nu} d$.}
\end{center}
\end{figure}

Now we show the explicit form of the coefficient functions 
$S_{LR}^l$ and $T_{LR}^l$ in the model. 
We calculate box diagrams, in which the left handed $W_{L}$ boson and 
the right handed gauge boson $W_R$ are exchanged, as seen in Fig.7. 
There are corresponding charged Higgs diagrams due to the gauge 
invariance. The internal upper fermion lines correspond to the ordinary
and the singlet quarks, the lower ones correspond to the SM
and the  singlet leptons.
\begin{figure}[H]
\begin{center}
\leavevmode\psfig{file=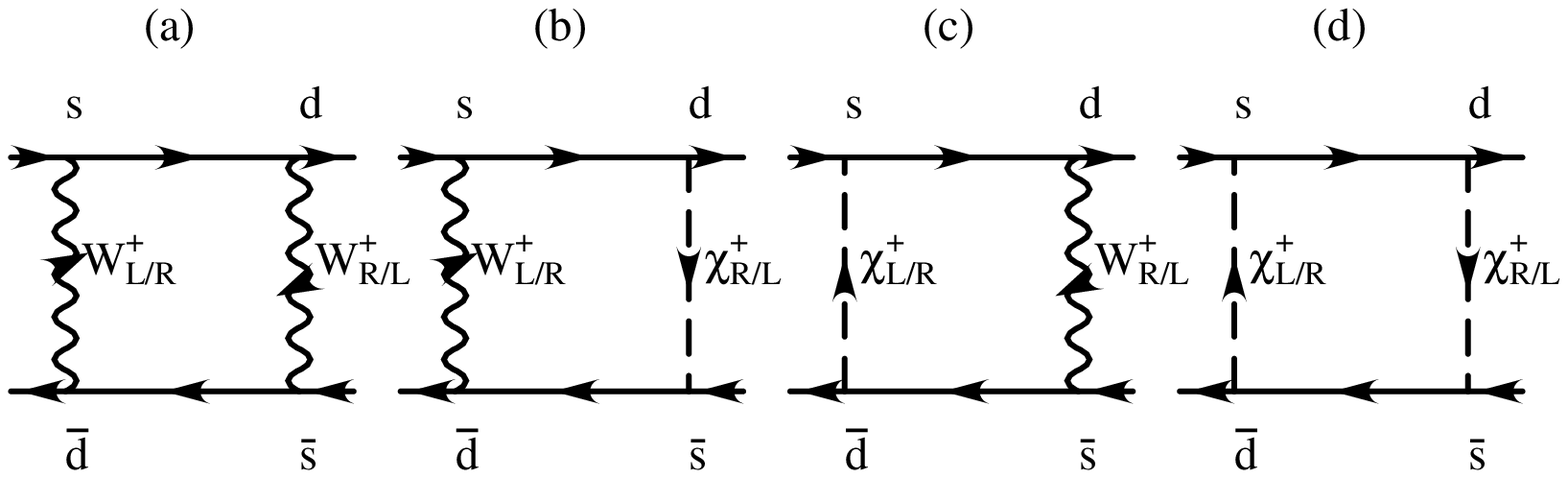,width=16cm}
\caption{Box diagrams which contribute to the effective 
Lagrangian for the process
$K \rightarrow \pi \nu \bar{\nu}$. 
(a) is a contribution from $W_{L}$ and $W_{R}$. (b) and (c) are gauge
boson and unphysical Higgs contributions. (d) is a contribution from the
unphysical Higgs $\chi_{L}$ and $\chi_{R}$.}
\end{center}
\end{figure}

The coefficients in the effective Lagrangian are:
\begin{eqnarray}
S^l_{LR}
&=&
c_{Ls} c_{Rd}\sum_{q=u,c,t}
(U^{L\ast}_{qs} U^{R}_{qd}) 
c_{Lq} c_{Rq} 
\times
\beta \sqrt{x_q y_l}
\tilde{F}\left(x_q,y_l,X_Q,Y_L,\beta \right),
\label{scalarcoe}
\\
T^l_{LR}
&=&
-
c_{Ls} c_{Rd}\sum_{q=u,c,t}
(U^{L\ast}_{qs} U^{R}_{qd}) c_{Lq} c_{Rq} 
\times
\beta \sqrt{x_q y_l}
\tilde{F}^{\prime}\left(x_q,y_l,X_Q,Y_L,\beta \right).
\label{tensorcoe}
\end{eqnarray}
where $\theta_{L/Rq}$ is the mixing angle between a singlet left/right
handed quark and the corresponding doublet quark
as defined in Eq.(\ref{Qqmix}), and  
$U^{L/R}_{ij}$ are 
$3\times 3$ CKM matrix elements. 
In addition to the parameters $x_{q}, X_{Q}$
and $\beta$ of Eq.(\ref{parameter}), 
new dimensionless parameters $y_{l},Y_{L}$  are defined by:
\begin{equation}
y_{l}=\frac{m_{l}^{2}}{M_{{W_L}}^{2}}, \hspace{1cm}
Y_{L}=\frac{m_{L}^{2}}{M_{{W_L}}^{2}}, \hspace{1cm}
\end{equation}
\noindent where $m_l$ are mass eigenvalues of the ordinary leptons $e$,
$\mu$ and 
$\tau$, while $m_L$ are the corresponding additional heavy lepton masses.

The function $\tilde{F}^{~\prime}$ is defined by replacing 
$F(x,y,\beta)$ with $F^{~\prime}(x,y,\beta)$ in Eq.(\ref{ftilde}):

\begin{eqnarray}
F^{\prime}(x,y,\beta)
&=&
\frac{1}{4}I_1 \left(x,y,\beta\right)
-
\frac{1+\beta}{16} I_2\left(x,y,\beta\right).
\end{eqnarray}
The other coefficients, $S_{RL}^l$ and $T_{RL}^l$, can be 
obtained by interchanging the indices $L$ and $R$. 
In the limit when $c_{L,s} = c_{R,d}=1$, 
$S_{LR}^l = S_{RL}^l$ and $T_{LR}^l = T_{RL}^l$. Thus, we simply write 
them as $S^l$ or $T^l$ and take the coefficients as LR symmetric. 
The coefficients $S^l$ ($l=e, \mu$) for electron and muon are
negligibly small due to the smallness of $y_{e}$ and $y_{\mu}$, 
and only $S^{\tau}$ contributes to the process significantly. 
The coefficient function $S^{\tau}$ evaluated with $M_{R}=500$ GeV is:
\begin{eqnarray}
S^\tau 
\cong 
-6.10 \times 10^{-7}\lambda_{u} 
-8.76 \times 10^{-5}\lambda_{c} 
-1.46 \times 10^{-3}c_{Rt} \lambda_{t}, 
\end{eqnarray}
where 
$\lambda_{u}\sim \lambda \left(1-\frac{\lambda^{3}}{2}\right)$, 
$\lambda_{c}\sim -\lambda \left(1-\frac{\lambda^{3}}{2}\right)$ and 
$\lambda_{t}\sim -A^2 \lambda^5 \left(1-\rho -i \eta\right)$ in 
the Wolfenstein parameterization. 
The coefficient for the up quark is negligible. 
For the top quark, the coefficient function is enhanced by its heavy mass,
but 
suppressed by the CKM factor $\lambda_{t}$, as
 compared to $\lambda_{c}$ from the 
charm quark.
Thus, the contribution from the top quark in the model is small in the case
of LR symmetric case. It is important that all the coefficients of 
the LR model are suppressed by the factor $\beta$, 
so the V-A interaction of the 
SM dominates as $M_{R}$ becomes large. 
\subsection{Pion energy spectrum}
We present the pion energy spectrum obtained
 using the coefficients of 
Eq.(\ref{scalarcoe}) and Eq.(\ref{tensorcoe}). In the process 
$K_L \rightarrow \pi^0 \nu \bar{\nu}$, the contributions from scalar 
interactions are controlled by $\beta$ and the enhancement factor 
$M_K/m_s$ in the matrix element $\langle (\bar{s}d) \rangle$.
For SM, the decay amplitude of $K_L \rightarrow \pi^0 \nu \bar{\nu}$
is proportional to $\eta$, thus the contributions of scalar interactions 
relative to that of SM for the pion energy spectrum is 
$ \rm{O}\left(\frac{M_K}{m_s}\frac{\beta}{\eta}\right)^2$. 
In the process $K^{+} \rightarrow \pi^{+} \nu \bar{\nu}$, contributions
from the scalar and tensor interactions are tiny compared to
the SM one.
The pion energy spectrum for the decay $K^{+} \rightarrow \pi^{+} \nu
\bar{\nu}$ is:
\begin{eqnarray}
\frac{dB\left(K^{+} \rightarrow \pi^+ \nu_{l} \bar{\nu}_{l}\right)}{d
x_{\pi}}
&=&
\delta_{N} \cdot \kappa_{+}\sqrt{x_{\pi}^2-4 \delta^2}
\nonumber \\
& & \hspace*{-5cm} \times
\left[ \left(x_{\pi}^2 -4 \delta^2\right)
       \left(| C_{SM}^l|^2 + 16 \omega_{+}^{2} \hat{t}~ |T^l|^2 \right)
+
3 \hat{t} \left(\frac{M_{K}}{m_{d}-m_{s}}\right)^2 
\left(1-\delta^2+\xi ~\hat{t} \right)^{2}
| S^l|^2 
\right] , 
\\
\kappa_+
&=&
\frac{3 \alpha^2 B(K^+ \rightarrow \pi^0 e^+ \nu) 
     }{2 \pi^2 \sin^4 \theta_{W} }
\lambda^8
=4.57 \times 10^{-11},
\\
\delta_{N}
&=&
\frac{ | f^+(K^+ \rightarrow \pi^0 e^+ \nu)|^2
      }{ \int d x_{\pi} 
         | f^+(K^+ \rightarrow \pi^0 e^+ \nu)|^2
         \left(x_{\pi}^2-4 \delta^2\right)^{3/2}
       },
\end{eqnarray}	
where $x_{\pi}$ is a normalized pion energy defined by $x_{\pi}=2 
E_{\pi}/M_{K}$, and $\delta=m_{\pi}/M_{K}$,
$\xi=f_{-}^{K^+\pi^0}/f_{+}^{K^+\pi^0}$,
$\omega_{+}=|f_{T}^{K^+ \pi^+}/f_{+}^{K^+\pi^0}|$ 
and $\hat{t}=(1+\delta^2-x_{\pi})$.

The process
$K_{L} \rightarrow \pi^0 \nu \bar{\nu}$ is a more sensitive probe of the 
scalar interactions, as discussed in the previous section. 
The energy spectrum is given by:
\begin{eqnarray}
\frac{dB\left(K_{L/S} \rightarrow \pi^0 \nu_{l} \bar{\nu}_{l}\right)}{d
x_{\pi}}
&=& \delta_{N} \cdot \kappa_{L}\sqrt{x_{\pi}^2-4 \delta^2}
\nonumber \\
\times && \hspace{-1cm}
\left[ 
(x_{\pi}^2 -4 \delta^2)
\left(~ |p C_{SM}^l \mp q C_{SM}^{l \ast}|^2 
      + \omega_{0}^2~\hat{t}~|p T^l \mp q T^{l \ast}|^2 ~
\right)
\right.
\nonumber \\
&&
\left.
+ 
3 \hat{t}\left(\frac{M_{K}}{m_{d}-m_{s}}\right)^2 
\left(1-\delta^2+\xi ~\hat{t}\right)^2 
|p S^l \pm q S^{l \ast} |^2 
\right],
\end{eqnarray}
where $\omega_0 \equiv |f_T^{K^0 \pi^0} /f_{+}^{K^+ \pi^0}|$ 
and $\kappa_L=\kappa_{+} \cdot \frac{\tau(K_L)}{\tau(K_+)}$.
The first term is contributions from V-A and tensor operators, 
$\bar{s}_{L,R}\sigma^{\mu\nu}d_{L,R}$, the second is the contribution from
the scalar operator 
which has an enhancement factor $ M_{K}^2/(m_{d}-m_{s})^2$ 
in the matrix element. The contribution from the tensor operator is
suppressed
by the kinematical factor $(x_{\pi}^{2}-4 \delta^2)$ in the low pion energy
region, 
$x_{\pi}\sim 2 \delta$, whereas at large pion energies, $x_{\pi}\sim 1+
\delta^2$, 
it is suppressed by the factor $\hat{t}$. Furthermore, decay through the
tensors is $CP$ violating and
has no enhancement factor. Hence, the contribution from the tensor operator
is negligible
compared to the scalar operator and the SM contributions.
The form factors are related to those of the experimentally well 
known decay mode $K^+ \rightarrow \pi^0 e^+\nu$.
In Fig.12 we show the pion energy spectrum, $\frac{d B(x_{\pi})}{d
x_{\pi}}$,
multiplied by a factor $10^{10}$.
\begin{figure}[H]
\begin{center}
\leavevmode\psfig{file=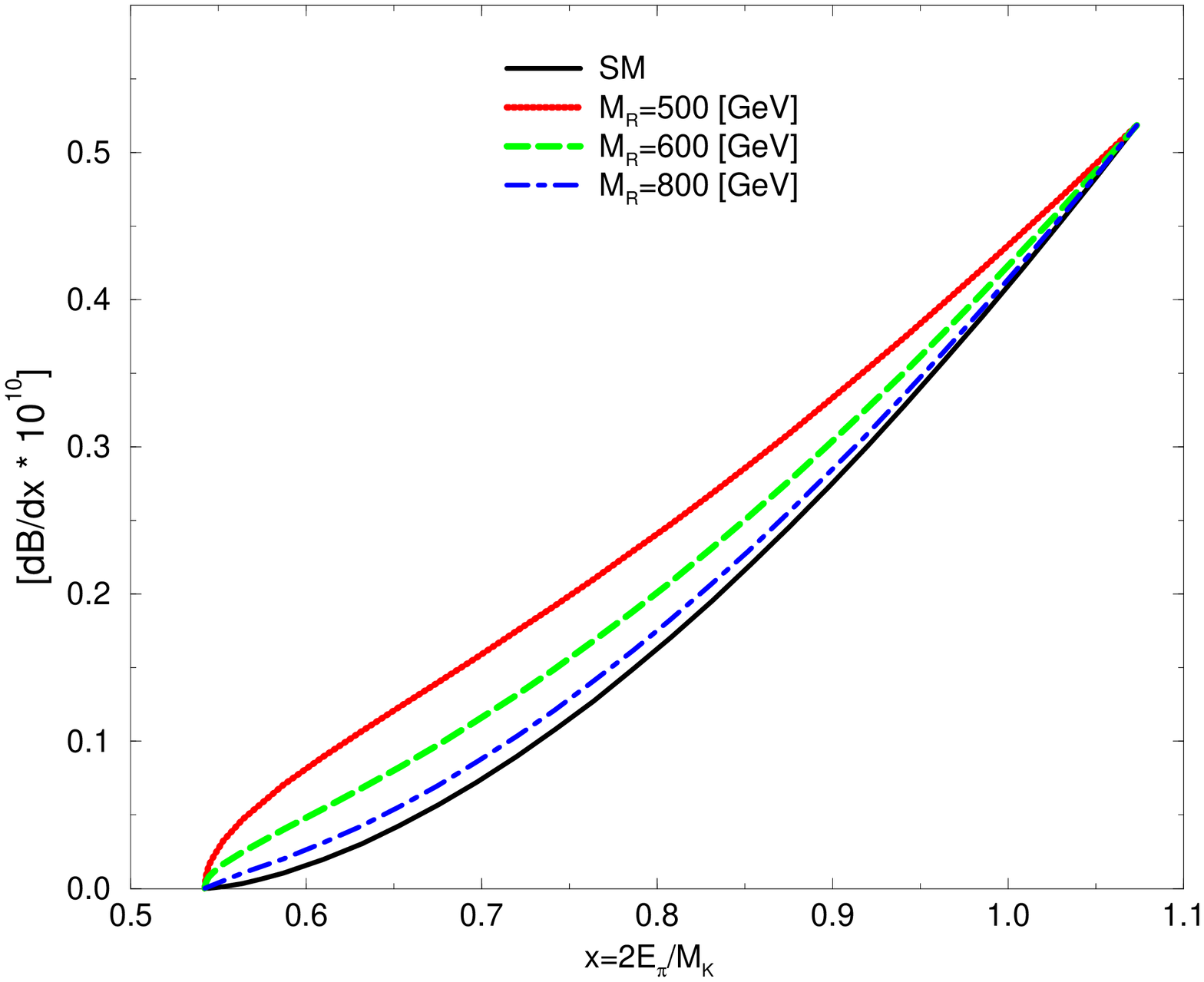,width=14cm,angle=-90}
\caption{The pion energy spectrum $
dB[K_L \rightarrow \pi^0 \nu {\bar \nu}]/dx $ for various values
for the 
$CP$ violating parameter  $\eta$. We use $m_s=100 (\rm MeV)$,
$M_{WR}=500 \rm(GeV)$ and $\rho=0.25.$
 SM denotes the predictions of the
standard model and LR denotes the predictions of the left-right
model.
}
\end{center}
\end{figure}
To see the dependence of the branching ratio on $\eta$, we plotted three
curves, 
which correspond to the cases $\eta=0.25, 0.3, 0.35$.
The solid lines are the pion energy spectra in the LR model, the
doted lines are the corresponding SM prediction, 
where we have taken $M_{W_{R}}=500 (\rm GeV)$ and $\rho=0.25$. 
The LR contribution is large in the low energy region 
$x_{\pi} \sim 2 \delta$, while in the high energy 
region the SM contribution dominates.  
In Fig.13, we study the dependence of the pion energy spectrum on
$M_{WR}$. 
Notice that as the coefficient functions of the effective Lagrangian are 
proportional to $\beta$, $d B/d x_{\pi}$ is proportional to 
$(M_{W_{L}}/M_{W_{R}})^4$. The effect is negligible for $M_{WR}>1 (\rm 
TeV)$ and the energy spectrum reduces to the prediction of the SM.

The effect of the new physics can be seen
 in the pion energy spectrum of the
$K_{L} \rightarrow \pi^0 \nu_{l} \bar{\nu}_{l}$ decay for
$M_{W_R} \leq 1\ {\rm TeV}$.
On the other hand, the analyses of $K^0-\bar K^0$ mixing
have given a constraint of $M_{W_R} \geq 1.6 \ {\rm TeV}$,
 which may discourage
the search for new physics in this decay mode.
However, it is important to comment on the constraint $M_{W_R} \geq
1.6 \ {\rm TeV}$.
This bound has been obtained 
from the box diagram of the $W_L-W_R$ intermediate state.
We assumed, as is usual, that the short
 distance effect dominates 
$K^0-\bar K^0$ mixing.
However, if long distance 
physics dominates the mixing, the constraint
of $M_{W_R} \geq 1.6 \ {\rm TeV}$ is no longer valid. 
\begin{figure}[H]
\begin{center}
\leavevmode\psfig{file=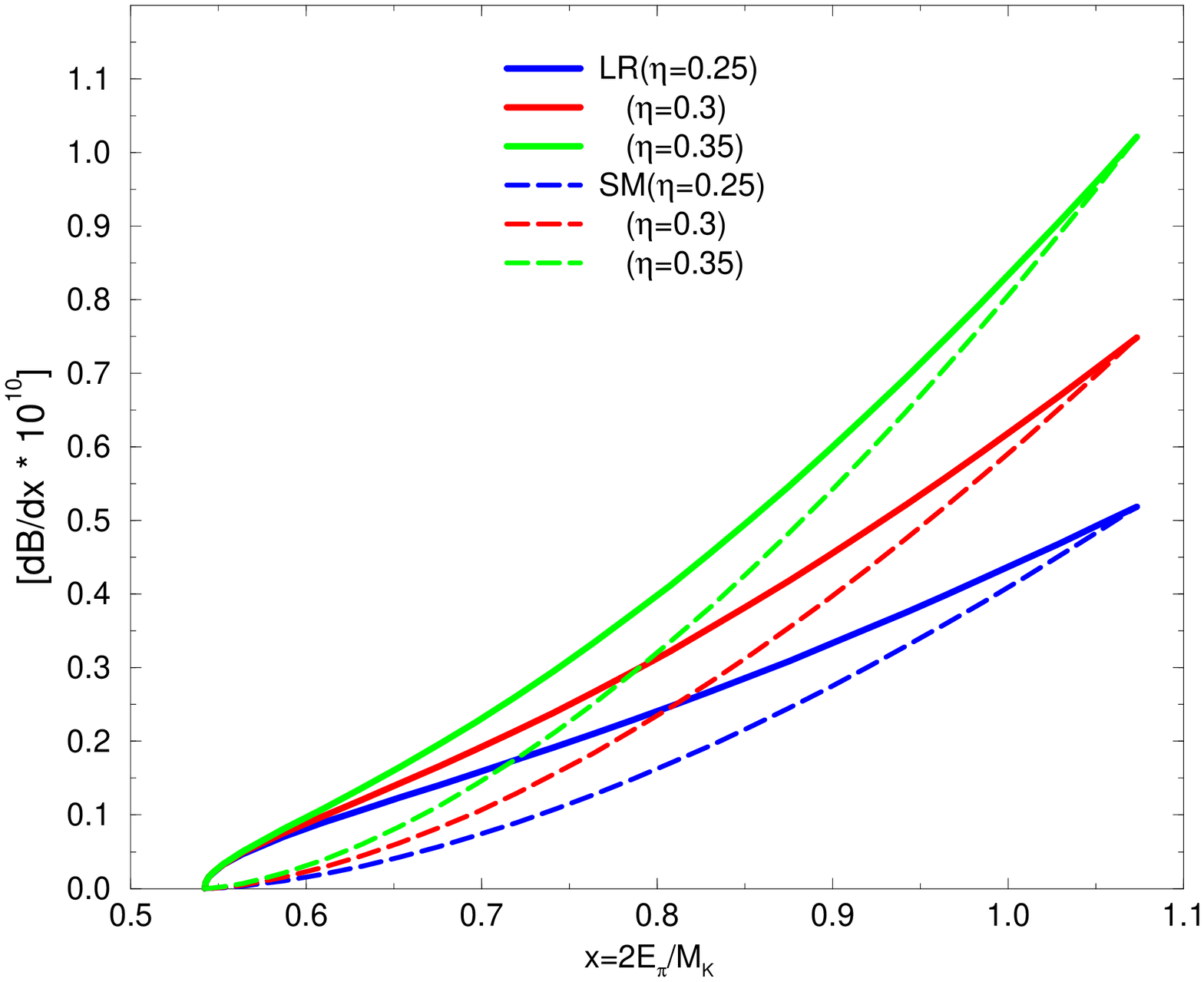,width=14cm,angle=-90}
\caption{ 
The pion energy spectrum $dB[K_L \rightarrow \pi^0 \nu {\bar\nu}]/dx$ 
for various values of $M_{W_R}$.
$m_s=100(\rm MeV)$ is used. SM denotes the prediction of the
standard model and LR denotes the predictions of the left-right
model.}
\end{center}
\end{figure}
\section{Conclusion}
We present the formalism and systematic analyses of the seesaw model for
quark masses. A framework allowing the top quark mass
to be of the order of the electroweak symmetry breaking scale
is explained. The derivation of the
quark mass formulae is presented in detail
with flavour mixing included.
There, it is shown that expanding simply by the
inverse power of the singlet
quark mass matrix fails and we propose an alternative
expansion to overcome the problem.
Furthermore, we find that a quark basis in which  the singlet and
doublet mixing Yukawa coupling is a triangular matrix
is appropriate for finding the mass base.
Starting in such a basis, by neglecting the flavor off-diagonal
Yukawa couplings, we can reproduce the quark mass formulae
which are obtained as the solutions of the
eigenvalue equation.
Also, we give the theoretical formulae
for the tree level FCNC.
 The tree level FCNC in the model are naturally
suppressed as ({\em quark masses})$^2$
{\em divided by an} $(SU(2)$ {\em breaking scale}$)^2$.
The effect on rare $K$ and $B$ decays is far below both the present
experimental bound and the prediction of the standard model.
As for FCNC beyond the tree level,
the one loop effect involving the right-handed gauge boson
exchange is  discussed
for  $K^0 - \bar K^0 $ mixing and for  $K \rightarrow \pi \nu \bar \nu$.
$\Delta m_K$ and $\epsilon_K$ 
give constraints on the parameters of the model $M_T$ and $M_{WR}$. 
We reanalyse the Beall, Bander and Soni bound for $M_{WR}$
in the present model and show  
the lower bound of $M_{WR}$ is about  $O (1\rm{TeV})$.
The constraint of $\epsilon_K$ is interesting and we 
show the allowed region in ($M_{WR}, {M_{T} \over M_{WR}}$).
If the allowed region of ($\rho, \eta$) is tightened 
by the data of the B factory and/or by the improvement of
the lattice computation of the hadronic matrix element, 
the allowed region 
will be much more specified than the present one. 
Alternatively   
the effect of the new physics can
 be seen  
the $K_{L} \rightarrow \pi^0 \nu_{l} \bar{\nu}_{l}$ decay 
in the case of $M_{WR} \le 1 ({\rm TeV}) $.
There is a scalar operator in the effective 
Lagrangian, which come from LR box diagrams.
For the decay 
 $K_{L}\rightarrow \pi^0 \nu \bar{\nu}$, there is a 
significant contribution from the scalar operator,
 especially in the low energy 
region of the pion energy spectrum, which, for the values 
$M_{W_{R}}=500 \G$ and $\rho,\eta=0.25$, amounts to an
enhancement of about $30\%$ to the total branching ratio.
Thus, measuring the decay $K_{L}\rightarrow \pi^0 \nu \bar{\nu}$ 
precisely may be  important to probe the effect from new physics.
Other aspects of the present model,
such as the constraints from the precision measurements and 
the other B decays,
the Higgs sector, and the neutrino mass and mixings
 will be discussed elsewhere.
The further improvement of the QCD corrections to our computation
is also needed for serious comparison with the experiments. 
\section*{\bf Acknowledgments} 
We  would like to thank  G. C. Branco for  discussions.
One of us (M.T) is also thankful to the High Energy Group 
in CFIF/IST for their hospitality.
This work is supported by the
Grant-in-Aid for Joint International Scientific Research
($\sharp 08044089$, Origin of $CP$ and $T$ violation and flavor physics) 
and the work of T.M. is supported in part by Grant-in-Aid for
Scientific Research on Priority Areas (Physics of $CP$ violation).

\end{document}